\documentclass[11pt,a4paper]{article}
\usepackage[utf8]{inputenc}
\usepackage[british]{babel}
\usepackage{amsmath,amssymb,amsfonts,bm}
\usepackage[margin=2.4cm]{geometry}
\usepackage{graphicx,float}
\usepackage{array,longtable}
\usepackage[font=small,labelfont=bf]{caption}

\usepackage{cite}
\usepackage{xcolor}
\definecolor{linkcol}{RGB}{30,70,120}
\usepackage[colorlinks=true,allcolors=linkcol]{hyperref}
\newcommand{\K}{\mathbf{K}}
\newcommand{\R}{\mathbf{R}}
\newcommand{\avg}[1]{\left\langle #1 \right\rangle}
\newcommand{\az}[1]{[#1]}

\title{\textbf{Hop rates of H and D on Pt(111) at finite coverage:\\ the memory equation and the collective friction}}
\author{S. Miret-Art\'es\\[4pt]
\small Instituto de F\'isica Fundamental, CSIC, Serrano 123, 28006 Madrid, Spain}
\date{\today}

\begin{document}
\maketitle

\begin{abstract}
\noindent
Helium scattering measures the coherent intermediate scattering function of an adlayer,
the characteristic function of the separation between adparticles at two times. Published dephasing rates of H and D on Pt(111) are analysed with its memory equation and
a nearest-neighbour Ising lattice gas. The measured rate belongs to the
adlayer. The hop rate of a single adsorbate is extracted from it by dividing out an
interaction factor, without a model of the barrier. This factor lies between $0.97$ and $1.17$ at $0.1$~ML, so that the published
single-adsorbate analyses hold there. At $0.4$~ML and $80$~K it lies between $1.2$ and
$4.7$, depending on how the occupied neighbours modify a tunnelling jump. The memory function is interpreted as a
friction, the collective friction, which the adlayer exerts on its own density
fluctuations. The lateral coupling, fixed by the coverage series, is the free energy of a
neighbouring pair, with a vibrational pair entropy. The small deformation of the measured curves from the Chudley--Elliott form, the tilt, is
proposed as an observable of the coupling. The tilts at $0.15$ and $0.2$~ML are consistent
with it, although they are hard to extract from the published curves. The memory equation
provides a transition-state theory for the adlayer, with the collective friction in the
role of the friction of the substrate and without divergence at the crossover. Combined
with the standard quantum transition-state theory of a single adsorbate, the extracted hop
rates give the friction of the substrate, $124$ and $75$~ps$^{-1}$ for H and D, independent
of the coverage.
\end{abstract}

\section{Introduction}

Hydrogen and deuterium on Pt(111) are the benchmark of the surface diffusion of light adsorbates.
Two helium scattering techniques have measured their motion. Quasielastic helium atom
scattering (QHAS) covers $130$ to $250$~K and $0.05$ to $0.66$ monolayer
(ML)~\cite{Graham1999}. Helium spin-echo (HSE) extends the range down to $80$~K at
$0.1$~ML~\cite{Jardine2010}. Both techniques measure the coherent intermediate
scattering function (ISF) of the adlayer. The coherent ISF is a characteristic function
(CF) in the sense of probability theory, the CF of the separation between adparticles
at two times~\cite{TorresMiyares2026a,PaperI}. Its decay is characterized by one rate,
the dephasing rate. The dephasing rate is a property of the layer. In this work it is
called measured only when it comes from the experiments. When it comes from kinetic Monte
Carlo simulations it is called simulated, and when it comes from closed forms,
calculated. A quantity obtained from measured rates with the help of a model, such as a
hop rate, is called extracted. The dependence of the dephasing rate on the momentum transfer shows
nearest-neighbour (NN) jumps between fcc hollow sites. Its temperature dependence shows
quantum tunnelling well above the crossover temperature of about $65$~K estimated in
Ref.~\cite{Jardine2010}. Its coverage
dependence was attributed to a NN repulsion of about $8$~meV~\cite{Graham1999}.

All theoretical analyses of these data treat a single adsorbate. They compute the hop
rate of an isolated adsorbate, the rate at which it leaves its site. Standard quantum
transition-state theory (TST) obtains it from a barrier, the friction of the substrate
and quantum corrections~\cite{Jardine2010,Hanggi1990,Wolynes1981}. The Grabert--Weiss
theory treats incoherent tunnelling between localized
states~\cite{Sanz2013,WeissGrabert1985,Grabert1985}. Stochastic wave functions of the
Lindblad equation solve the dissipative quantum dynamics of one
adsorbate~\cite{TorresMiyares2023,TorresMiyares2025}. The CF method describes the signal of a single adsorbate
as the CF of its displacement~\cite{TorresMiyares2026a}. All of them interpret the measured
rate as that of an isolated adsorbate. Helium scattering, however, measures a coherent signal,
which correlates different adparticles. At $0.1$~ML an adparticle has on average $0.6$
occupied NN sites, and a repulsion of $8$~meV is comparable to the thermal energy
$k_BT$, with $k_B$ the Boltzmann constant and $T$ the surface temperature. The measured
dephasing rate is therefore a property of the adlayer, not of an isolated adsorbate. The
adlayer does not alter the measurement. It alters its meaning. Taken as a single-adsorbate
rate, the measured rate can give wrong physical parameters of the single-adsorbate
diffusion, even at the lowest coverage studied. The question is how large this error is
and how to correct it.

The memory equation of the coherent ISF answers this question for a class of lattice
gases~\cite{PaperI}. It holds when the motion obeys detailed balance. Incoherent
tunnelling belongs to this class, since each adsorbate loses its quantum phase between
transitions and the bath is in equilibrium. The equation was applied to the Ising
lattice gas and to CO/Cu(001), where the lateral coupling and the memory of the layer
were obtained from the dephasing rates measured on two azimuths~\cite{PaperII}. It
separates the single-adsorbate hop rate, which contains all the quantum physics and the
friction of the substrate, from the effects of the adlayer.

The resulting description can be seen as a TST of the adlayer. It is complementary to
the single-adsorbate theories. Whatever theory gives the hop rate of an isolated adsorbate,
the adlayer multiplies it by an interaction factor $g$. Unlike the TST of one adsorbate, it
has no divergence at the crossover, and it holds for incoherent tunnelling at every
temperature. The memory function of the equation is interpreted here as a friction, called the
collective, or configurational, friction. It is exerted
by the adlayer on its own density fluctuations. Its integral fixes the measure of the
memory used in Refs.~\cite{PaperI,PaperII}, which therefore quantifies this friction. The friction of the substrate acts
instead on each adsorbate. The two act on different variables, so that they cannot be added
into a total friction. The collisional friction of the interacting single-adsorbate
approximation was an initial and simple attempt to include the effect of the other
adsorbates as a friction on one adsorbate~\cite{MartinezCasado2007,MartinezCasado2008}.
Simulations show that the collective friction has a strong memory. It nearly vanishes
at long wavelength, where the CF method gives an Einstein law for collective diffusion.

The aim of the analysis is twofold. The first aim is to translate the dephasing rates
measured at finite coverage into the hop rate of a single adsorbate. The physics of the
isolated adsorbate lies in this hop rate. It contains the barrier, the quantum corrections,
with the tunnelling and the crossover, and the friction of the substrate. These are what
the experiments are meant to reveal and what the single-adsorbate theories compute.
Helium scattering, however, needs a finite coverage to give a measurable signal, and the
rate measured at the lowest coverage is already a property of the adlayer. Only after the
effect of the adlayer is divided out can the barrier, the tunnelling and the friction of the substrate be
extracted from the data and compared with theory. The second aim is to obtain the
properties of the adlayer, its lateral coupling and its collective friction. The analysis extracts the single-adsorbate hop rate from the
measured rates without a model of the barrier. It uses their dependence on the momentum
transfer, on two azimuths, and on the coverage. For an isolated adsorbate the dephasing rate
against the momentum transfer is the Chudley--Elliott curve, whose form is fixed by
the lattice and whose height is set by the hop rate. The measured curves keep this form
up to a small deformation. Their height, divided by the interaction factor $g$, gives the
single-adsorbate hop rate. Their deformation from the Chudley--Elliott curve is quantified
by a new quantity, the tilt. A single-adsorbate analysis assumes the Chudley--Elliott
curve and so sets the tilt to zero. Within the lattice gas the tilt depends on the
lateral coupling alone. The coverage dependence of the rates
fixes the coupling, which turns out to be a free energy, with an entropy from the
softening of the vibrations by an occupied neighbour. The tilts give an independent but weaker test of this coupling, since they are hard to
extract from the published curves. The analysis thus applies both to rates measured at a single momentum
transfer, given a coverage series, and to momentum-resolved curves, which also carry the
tilt. The analysis also separates the coupling from the friction of the substrate, which is
recovered from the hop rates and turns out to be independent of the coverage. The hop
rates are finally compared with standard quantum TST and with stochastic wave functions.

The paper is organized in seven sections. Section~2 presents the memory equation, the
collective friction and the geometry of the two azimuths. Section~3 gives the statics
of the lattice gas, with the free-energy coupling and the static structure factor. Section~4
treats the dynamics, first of a single adsorbate and then of the adlayer, and defines
the interaction factor $g$, the tilt and the collective diffusion coefficient. Section~5
first describes the simulations and the properties of the adlayer found in them and then analyses
the measured rates. The coverage series fixes the coupling, and the
momentum-resolved curves give the hop rates and test the coupling through their tilts.
Section~6 compares the memory equation with standard quantum TST. It extracts the parameters of
the adsorbate--substrate system, presents the memory equation as a TST of the adlayer,
and combines the two into one rate of the adlayer.
Section~7 summarizes the conclusions. The Appendix collects simulation results, the least-squares fits, the data treatment and
additional analyses.

\section{The framework}

In a lattice gas each adsorption site $\R$ is empty or holds one adparticle, with the
occupation number $n_{\R}=0$ or $1$. The set of all occupation numbers is the
configuration of the adlayer, and their mean is the coverage $\theta$. The memory
equation holds for a class of lattice gases~\cite{PaperI}. In this class the adparticles
are conserved and jump to empty sites by fixed jump vectors. The jump rates may depend
on the configuration but obey detailed balance, and the dynamics is invariant under the
lattice translations and inversion. Two kinds of rate are distinguished, as in
Ref.~\cite{PaperII}. A jump rate refers to one jump vector $\mathbf L$. It is the rate at
which an adparticle moves by $\mathbf L$ in a given configuration. A hop rate refers to the
site. It is the rate at which an adparticle leaves its site by any jump vector, the sum
of its jump rates. The total hop rate $\Gamma_0$ is that of an isolated adparticle. The
effective hop rate $\Gamma_{\rm eff}$ is its mean over the adparticles of the layer.

The ISF $I(\K,t)$ depends on the momentum transfer $\K$ parallel to the surface and on
the time $t$. It is the correlation of the density wave, the Fourier component at $\K$
of the deviations of the occupation numbers from $\theta$,
\begin{equation}
\delta n_{\K}(t)\;=\;\sum_{\R}\bigl[n_{\R}(t)-\theta\bigr]e^{\,i\K\cdot\R} ,
\qquad
I(\K,t)\;=\;\frac{\avg{\delta n_{\K}^*(0)\,\delta n_{\K}(t)}}{N} ,
\label{eq:dw}
\end{equation}
with $\avg{\cdot}$ the equilibrium average over the configurations and $N=\theta N_s$ the
number of adparticles on $N_s$ sites~\cite{PaperII}. The ISF is a CF in the strict sense of
probability theory~\cite{TorresMiyares2026a,PaperI}. Let
$\rho_{\R}(t)=\avg{n_{\mathbf 0}(0)\,n_{\R}(t)}/\theta$ be the probability that the site
at separation $\R$ is occupied at time $t$, given that the origin was occupied at
$t=0$. Then
\begin{equation}
I(\K,t)\;=\;\sum_{\R}\bigl[\rho_{\R}(t)-\theta\bigr]e^{\,i\K\cdot\R} ,
\qquad
\rho_{\R}(t)-\theta\;=\;\frac1{N_s}\sum_{\K}I(\K,t)\,e^{-i\K\cdot\R} ,
\label{eq:cf}
\end{equation}
with the inverse sum over the first Brillouin zone. The
uniform part $\theta$ gives the Bragg peaks and is removed. The memory equation is the
equation of motion of this CF~\cite{PaperI}. The static
structure factor (SSF) is $F(\K)=I(\K,0)$. In the dilute limit $I/F$ is the CF of the
displacement of one adsorbate, the self ISF. At finite coverage the coherent ISF
correlates different adparticles.

The decay of the ISF passes through three regimes~\cite{PaperI,PaperII}. In the ballistic
regime, at the shortest times, each adparticle moves freely in its well, as a single
adsorbate does~\cite{Vineyard1958}. The friction of the substrate ends this regime within
$7$--$14$~fs for H on Pt(111)~\cite{TorresMiyares2025}, long before the first jumps. In the
jump regime, over times of the order of the residence time at a site, the adparticles
jump and the memory of earlier jumps acts. In the diffusive regime, at times much longer
than the residence time, the fast modes have died out and only the slowest relaxation
remains. At long wavelength its rate is proportional to the square of the momentum transfer. In the jump and diffusive
regimes the ISF decays as a sum of exponentials with positive weights~\cite{PaperI},
\begin{equation}
\frac{I(\K,t)}{F(\K)}\;=\;\sum_k a_k(\K)\,e^{-\lambda_k(\K)t} ,
\qquad
a_k(\K)>0 ,\quad \sum_ka_k(\K)=1 .
\label{eq:modes}
\end{equation}
Each exponential is a relaxation mode of the density wave, with the rate $\lambda_k$
and the weight $a_k$. The lowest rate is $\lambda_{\min}$. The decay starts at the
local rate $\Omega(\K)=\sum_ka_k\lambda_k$, fixed by an exact sum rule. The
correlation time, the area beneath $I/F$, is $\tau_c=\sum_ka_k/\lambda_k$. Hence
$\lambda_{\min}\le\tau_c^{-1}\le\Omega$. As in Refs.~\cite{PaperI,PaperII}, the memory is
quantified by $R(\K)=\Omega\tau_c\ge1$. It equals one when every region of the layer relaxes the
density wave at the same rate. Where neighbouring adparticles repel, the relaxation is
fast. Where several adparticles must move, it is slow. A hop rate obtained from the area
beneath the decay is then too small by the factor $R$.

Both experiments reduce each measured decay to one rate, the dephasing rate
$\alpha(\K)$. HSE measures the ISF in time and fits a single exponential $e^{-\alpha t}$
over a time window. The window is placed in the diffusive regime, where the fitted rate
changes only slowly. Only the slowest mode is left there, and the fit gives its
rate~\cite{PaperI,PaperII},
\begin{equation}
\alpha(\K)\;=\;\lambda_{\min}(\K)\;\le\;\tau_c^{-1}(\K)\;\le\;\Omega(\K) .
\label{eq:alphaR}
\end{equation}
The correlation time takes in the whole decay, the jump regime included. It is reached
only through the area beneath the decay, and $\tau_c^{-1}$ equals $\lambda_{\min}$ only when
one mode carries the whole weight. QHAS measures the dynamic structure factor $S(\K,\omega)$, the Fourier transform of the
ISF in time. It fits a Lorentzian to the quasielastic peak around $\omega=0$. Its full
width at half maximum is $\Gamma_{\rm QHAS}=2\hbar\alpha$, with $\hbar$ the reduced Planck
constant. The QHAS widths thus give the same dephasing rate $\alpha$.

The memory equation describes the same decay. The projection of Nakajima and Zwanzig~\cite{Nakajima1958,Zwanzig1960}
eliminates every variable of the layer except the density wave and leaves at every
momentum transfer an exact equation of Mori's form~\cite{PaperI,Mori1965},
\begin{equation}
\frac{\partial I(\K,t)}{\partial t}
\;=\;-\,\Omega(\K)\,I(\K,t)\;+\;\int_0^t dt'\,M(\K,t-t')\,I(\K,t') ,
\qquad I(\K,0)=F(\K) .
\label{eq:eom}
\end{equation}
Here $M(\K,\tau)$ is the memory function, or kernel, at the delay $\tau$. It is the correlation function
of the part of the flux not carried by the density wave itself. Equation~\eqref{eq:eom} holds in the jump and diffusive
regimes. It has the form of a Pauli master equation with memory, here for the density
wave. The initial value of the memory function, which
gives its strength, is the spread of the relaxation rates~\cite{PaperI},
\begin{equation}
M(\K,0)\;=\;m_2(\K)-\Omega^2(\K) ,
\qquad
m_2\;=\;\sum_ka_k\lambda_k^2 ,
\label{eq:M0}
\end{equation}
where $m_2$ is a static average of the layer, the correlation between jumps on two pairs
of sites~\cite{PaperI}. Let $\hat I(\K,s)=\int_0^\infty dt\,e^{-st}I(\K,t)$ be the Laplace
transform, so that $\hat I/F=\sum_ka_k/(s+\lambda_k)$. Equation~\eqref{eq:eom} becomes
\begin{equation}
\frac{\hat I(\K,s)}{F(\K)}\;=\;\frac{1}{s+\Omega(\K)-\hat M(\K,s)} ,
\label{eq:laplace}
\end{equation}
with $\hat M$ the Laplace transform of the memory function. Physically, a jump relaxes the density wave, and the correlated
back jumps partly restore it. The memory function describes this restoring flux and how long it
lasts. Equation~\eqref{eq:modes} fixes two properties of the memory function. On the negative
axis of $s$, between two consecutive relaxation rates, $\hat I/F$ decreases
monotonically from $+\infty$ to $-\infty$ and vanishes once. These zeros are the poles
of $\hat M$, one between each pair of consecutive relaxation rates, and the residues are
positive. The memory function is thus a sum of exponentials with positive weights, and all its
rates exceed $\lambda_{\min}$. A continuous spectrum
gives the same result~\cite{Schilling2012}. At $s=0$, $\tau_c^{-1}=\Omega-\hat M(\K,0)$, which gives the meaning of $R$ found in
Ref.~\cite{PaperI},
\begin{equation}
\frac{1}{R(\K)}\;=\;1-\frac{\hat M(\K,0)}{\Omega(\K)} ,
\qquad
\hat M(\K,0)\;=\;\int_0^\infty d\tau\,M(\K,\tau) .
\label{eq:R}
\end{equation}
The memory measure $R$ is thus fixed by the integral of the memory function over the delay, and
$1-1/R$ is the fraction of the local rate that the memory function removes.
At the first pole of $\hat I$,
\begin{equation}
\lambda_{\min}\;=\;\Omega(\K)-\hat M(\K,-\lambda_{\min}) .
\label{eq:pole}
\end{equation}
Since every rate of the memory function exceeds $\lambda_{\min}$,
$\hat M(\K,-\lambda_{\min})\ge\hat M(\K,0)\ge0$, and the bounds are recovered.

The memory function lowers the relaxation rate below the local rate $\Omega$, as a friction slows
a motion. It is therefore called the collective, or configurational, friction. It is
collective because it acts on the density wave, a collective variable. It is
configurational because its bath is the set of the configurations of the layer, through
the other relaxation modes. It is a friction although the dynamics has no momentum. It
is non-negative at every delay~\cite{PaperI}, it comes from a bath, and it lowers the
rate below its initial value, as the friction kernel does in Mori's equation for a
velocity~\cite{Mori1965}. Friction kernels with memory on collective coordinates are
a standard tool of nonequilibrium statistical mechanics~\cite{Zwanzig2001} and are now
extracted from simulations across soft matter~\cite{Dalton2024}. For the folding
coordinate of proteins, for instance, the memory of the friction changes the folding
rates~\cite{Dalton2023}. Here the term denotes the memory function of the coherent ISF of a
lattice gas, and the short form collective friction is used below.
Memory functions of the density in interacting lattice gases have been written as
continued fractions~\cite{Danani1997} and reviewed for collective surface
diffusion~\cite{AlaNissila2002}. There they are mostly approximated, as in dynamical
mean-field theory, or neglected, as in the Darken approximation. The present work
interprets the memory function as a friction, gives it in closed form, and applies it to
helium-scattering data. It contains no friction of the substrate, which acts on each
adsorbate and enters only through $\Gamma_0$. The two act on different variables and
cannot be combined into a total friction.

The local rate obeys an exact sum rule~\cite{PaperI,PaperII},
\begin{equation}
\Omega(\K)\,F(\K)\;=\;\Gamma_{\rm eff}\bigl[1-f_{\K}\bigr] ,
\qquad
J_{\mathbf L}\;=\;\frac1\theta\Bigl\langle n_{\R}\bigl(1-n_{\R+\mathbf L}\bigr)\,w_{\R,\mathbf L}\Bigr\rangle ,
\qquad
\Gamma_{\rm eff}\;=\;\sum_{\mathbf L}J_{\mathbf L} .
\label{eq:sumrule-gen}
\end{equation}
Here $w_{\R,\mathbf L}$ is the rate of the jump from $\R$ by $\mathbf L$ in the given
configuration. The factor $n_{\R}(1-n_{\R+\mathbf L})$ requires an occupied departure
site and an empty target, and the average runs over the equilibrium configurations.
$J_{\mathbf L}$ is thus the equilibrium number of jumps by $\mathbf L$ per unit time and
adparticle. $f_{\K}=\sum_{\mathbf L}p_{\mathbf L}e^{i\K\cdot\mathbf L}$ is the CF of one
jump vector, with $p_{\mathbf L}=J_{\mathbf L}/\Gamma_{\rm eff}$ the fraction of jumps made
by $\mathbf L$. A jump orthogonal to $\K$ drops out of the factor $1-f_{\K}$ but still acts through $\Gamma_{\rm eff}$, $F(\K)$ and the memory~\cite{PaperI}. For an
isolated adparticle $\Gamma_{\rm eff}=\Gamma_0$ and $F=1$, and the ISF is the
Chudley--Elliott decay $e^{-\Gamma_0[1-f_{\K}]t}$~\cite{PaperII}. The rates of the signal,
$\Omega$ and $\alpha$, are thus distinct from the jump and hop rates.

H,D/Pt(111) belongs to this class. Both experiments find NN jumps between fcc sites
alone~\cite{Jardine2010,Graham1999}. These sites form a triangular lattice of spacing
$a=2.77$~\AA. The coverage is
conserved, and the measured length scales lie on the terraces~\cite{Jardine2010}. Above
the crossover the tunnelling is incoherent. Each adsorbate loses its quantum phase between
transitions and moves by jumps governed by a master
equation~\cite{WeissGrabert1985,Sanz2013,TorresMiyares2025}. Incoherent refers to the
motion of each adsorbate. The scattering is coherent because it correlates different adsorbates.
For transitions coupled to a bath in equilibrium, incoherent tunnelling included, the
rates of a jump and of its reverse stand in the ratio $e^{-\beta\Delta H}$, with $\Delta
H$ the change of interaction energy and $\beta=1/k_BT$~\cite{Weiss1999}. This is
detailed balance. The six jump vectors form three pairs $\pm\mathbf L_j$, $j=1,2,3$, one
pair per symmetry direction. On the flat terrace they are equivalent, and each carries
the fraction $\tfrac16$ of the jumps. Let $\psi_j$ be the angle between $\K$ and
$\mathbf L_j$. Then $f_{\K}=\tfrac13\sum_j\cos(Ka\cos\psi_j)$, with $K$ the modulus of
$\K$, and the sum rule \eqref{eq:sumrule-gen} becomes
\begin{equation}
\Omega(\K)\,F(\K)\;=\;\frac{\Gamma_{\rm eff}}3\sum_{j=1}^{3}\bigl[1-\cos\bigl(Ka\cos\psi_j\bigr)\bigr]
\;=\;\Gamma_{\rm eff}\bigl[1-f_{\K}\bigr] .
\label{eq:sumrule}
\end{equation}
Each pair $\pm\mathbf L_j$ contributes the rate $\Gamma_{\rm eff}/3$, the sum of its two
jump rates.
The $\az{1\bar10}$ azimuth is parallel to $\mathbf L_3$, and $\az{11\bar2}$ is
orthogonal to it.

\begin{figure}[!t]
\centering
\includegraphics[width=0.99\textwidth]{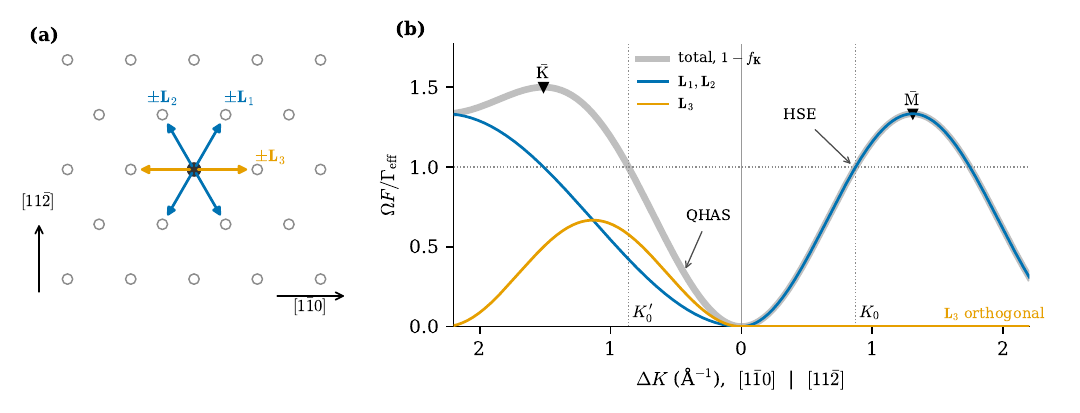}
\caption{Direction of the momentum transfer on Pt(111). (a) fcc sites (circles), the
jump vectors $\pm\mathbf L_1,\pm\mathbf L_2$ (blue) and $\pm\mathbf L_3$ (orange), and
the azimuths $\az{11\bar2}$ and $\az{1\bar10}$. (b) Right-hand side of the sum rule
\eqref{eq:sumrule} in units of $\Gamma_{\rm eff}$ on $\az{1\bar10}$ (left) and $\az{11\bar2}$
(right). Shown are the contributions of $\mathbf L_1,\mathbf L_2$ (blue) and
$\mathbf L_3$ (orange) and the total $1-f_{\K}$ (grey). Dotted lines mark the nodes
$K_0$, $K_0'$ of $f_{\K}$, where the total equals one. $\bar{\rm M}$ and $\bar{\rm K}$
are zone-boundary points. Arrows mark the momentum transfers of the QHAS measurements
of Ref.~\cite{Graham1999} and the HSE measurements of Ref.~\cite{Jardine2010}. All
curves are calculated from the lattice geometry. Nothing is fitted.}
\label{fig:directions}
\end{figure}

Figure~\ref{fig:directions} shows which jumps each azimuth resolves. Panel (a) displays
the sites, the jump directions and the azimuths. Panel (b) displays the right-hand side
of \eqref{eq:sumrule} on both azimuths, with the contributions of each direction and
the momentum transfers of the experiments. On $\az{11\bar2}$, $\mathbf L_1$ and
$\mathbf L_2$ form the angle $\pi/6$ with $\K$, and $\mathbf L_3$ is orthogonal to it.
The right-hand side is $\tfrac23\Gamma_{\rm eff}[1-\cos(Ka\cos\tfrac{\pi}{6})]$, the single cosine of
the CF analysis~\cite{TorresMiyares2026a}, up to the zone-boundary point $\bar{\rm M}$,
$K=1.31$~\AA$^{-1}$, where $f_{\K}=-\tfrac13$. On $\az{1\bar10}$ all three directions
act, $\mathbf L_3$ with its full length and $\mathbf L_1$, $\mathbf L_2$ with half of it.
The sum reaches $1.5\,\Gamma_{\rm eff}$ at the zone corner $\bar{\rm K}$, $K=1.51$~\AA$^{-1}$, where
$f_{\K}=-\tfrac12$. On both azimuths $f_{\K}$ vanishes near $0.87$~\AA$^{-1}$, at
$K_0=4\pi/(3\sqrt3a)=0.873$~\AA$^{-1}$ on $\az{11\bar2}$ and $K_0'=0.864$~\AA$^{-1}$ on
$\az{1\bar10}$. At these nodes the resolved factor $1-f_{\K}$ is one on both azimuths. The HSE
Arrhenius plot was measured practically on the node, at $0.86$~\AA$^{-1}$ on
$\az{11\bar2}$~\cite{Jardine2010}. The QHAS coverage series was measured at
$0.44$~\AA$^{-1}$ on $\az{1\bar10}$, where $f_{\K}=0.66$~\cite{Graham1999}.

\section{Statics of the lattice gas}

This section gives the equilibrium quantities of the layer. They fix the static part
of the effect of the layer on the dephasing rate. They depend on the coupling, defined below, only
through the reduced coupling $\beta\varepsilon$ and not on the hop rate or on either
friction. The collective friction is dynamical and does not appear here. In the NN Ising lattice gas the energy
of a configuration is
\begin{equation}
H\;=\;-\varepsilon\sum_{\langle\R,\R'\rangle}n_{\R}\,n_{\R'} ,
\label{eq:H}
\end{equation}
where the sum runs over all pairs of NN sites. Two adparticles on NN sites thus form a
bond of energy $-\varepsilon$, and $\varepsilon<0$ is a repulsion. A real adsorbate also
vibrates in its well, and an occupied neighbour can change both the energy and the shape
of that well. The coupling $\varepsilon$ is therefore obtained by summing over the
vibrations at fixed occupations, the standard route from a lattice model with internal
degrees of freedom to an Ising model~\cite{Hill1960,FowlerGuggenheim1939}.

An isolated adparticle has three vibrational modes, two parallel and one normal to the
surface. Near the bottom of the well they are independent harmonic oscillators. The
threefold symmetry of the fcc site makes the two parallel modes degenerate. The modes
are also nearly independent of the substrate phonons, since the mass of H is about
$1/195$ of that of Pt. Anharmonic couplings are neglected. With $\omega_k$, $k=1,2,3$,
the mode frequencies, the vibrational partition function of the adparticle is
\begin{equation}
Q_0\;=\;\prod_{k=1}^{3}Q(\omega_k) ,
\qquad
Q(\omega)\;=\;\sum_{n=0}^{\infty}e^{-\beta\hbar\omega(n+1/2)}
\;=\;\frac{1}{2\sinh(\beta\hbar\omega/2)} .
\label{eq:Q0}
\end{equation}
The energies are counted from the bottom of the well, so that $Q$ includes the
zero-point energy. In the classical limit, $k_BT\gg\hbar\omega$,
$Q(\omega)=k_BT/\hbar\omega$.

Two adparticles on NN sites differ from two isolated ones in two ways. The bottom of
each well is higher by the pair energy $e_p$, and the frequencies of the three modes
of each adparticle change from $\omega_k$ to $\omega_k'$. Let both changes be additive over the occupied
neighbours. Each NN pair then multiplies the partition function by $e^{-\beta e_p}$ and
by $\prod_kQ(\omega_k')/Q(\omega_k)$, which defines the vibrational free energy of the
pair,
\begin{equation}
\Delta f_{\rm vib}(T)\;=\;-k_BT\sum_k\ln\frac{Q(\omega_k')}{Q(\omega_k)} .
\label{eq:fvib}
\end{equation}
The canonical partition function of $N$ adparticles on $N_s$ sites is then
\begin{equation}
Z\;=\;Q_0^{N}\sum_{\{n_{\R}\}}e^{-\beta\,[e_p+\Delta f_{\rm vib}(T)]\,N_{\rm nn}} ,
\label{eq:Z}
\end{equation}
with $N_{\rm nn}$ the number of NN pairs and the sum over all configurations with $N$
adparticles. The factor $Q_0^N$ only shifts the chemical potential and affects neither
the structure nor the dynamics at fixed coverage. Equation~\eqref{eq:Z} is thus the
partition function of the lattice gas \eqref{eq:H} with
$-\varepsilon=e_p+\Delta f_{\rm vib}(T)$. The coupling is the free energy of forming one
NN pair, not simply an energy.

The vibrational free energy has an energy part and an entropy part $-Ts_p$. The pair
entropy $s_p$ is the change of vibrational entropy of the two adparticles when they
become neighbours. In the classical limit the vibrational energy of a mode is $k_BT$
whatever the frequency, so that the energy part vanishes and
\begin{equation}
s_p\;=\;k_B\sum_k\ln\frac{\omega_k}{\omega_k'} ,
\label{eq:sp}
\end{equation}
independent of temperature. For quantum modes a small energy change remains, mostly
from the zero-point energy. It is absorbed into $e_p$, and both terms are taken constant
over the range of a fit. A neighbour that lowers the curvature of the well, a softening
of the well, lowers its frequencies, and $s_p>0$. The coupling is
\begin{equation}
-\varepsilon(T)\;=\;e_p-Ts_p .
\label{eq:fp}
\end{equation}
The pair energy makes a NN pair costly, and the pair entropy makes it favourable, more
so as the temperature rises. The pair entropy is vibrational and differs from the configurational entropy, which
is generated by the sum in \eqref{eq:Z}. A friction of the substrate can also lower the
prefactor of a rate, but only through the hop rate of each adparticle. The pair entropy
instead changes the coupling and hence the structure of the layer.

Without softening, $s_p=0$ and the coupling is the constant $\varepsilon=-e_p$. With
softening, $s_p>0$ and
\begin{equation}
\varepsilon(T)\;=\;-e_p\Bigl(1-\frac{T}{T^*}\Bigr) ,
\qquad
T^*=\frac{e_p}{s_p} .
\label{eq:epsT}
\end{equation}
$T^*$ is called the compensation temperature. At $T^*$ the energy cost of a pair equals its
entropy gain, and the coupling vanishes. Below $T^*$ the adparticles repel, more weakly as
the temperature rises. At $T^*$ the layer is ideal at every coverage. Above $T^*$ the
coupling turns into a weak attraction. The equilibrium depends on the coupling only
through the reduced coupling $\beta\varepsilon$. By \eqref{eq:epsT},
$\beta\varepsilon=-(e_p/k_B)(1/T-1/T^*)$. For a constant coupling the reduced coupling
grows as $1/T$. For $\varepsilon(T)$ it passes through zero at $T^*$. The name comes from the compensation, or Meyer--Neldel,
rule~\cite{MeyerNeldel1937,Yelon1992}. Consider a family of rates, here at different
coverages, each obeying an Arrhenius law $\nu_0e^{-E_d/k_BT}$, with $E_d$ the activation
energy and $\nu_0$ the prefactor. If a change within the
family shifts the activation energy and the logarithm of the prefactor in the fixed
ratio $\Delta E_d=k_BT^*\Delta\ln\nu_0$, all Arrhenius lines cross at the compensation
temperature $T^*$, where all rates coincide. The coupling \eqref{eq:epsT} produces such
a family. Its energy part shifts the activation energy, and its entropy part the
prefactor, both in proportion to the number of bonds. Compensation is common in surface
diffusion and other activated processes~\cite{Yelon1992}.

The structure of the layer is obtained from the pair approximation and the random-phase
approximation (RPA)~\cite{PaperII}. The pair approximation, also called quasi-chemical or Bethe
approximation~\cite{FowlerGuggenheim1939,Hill1960}, neglects the correlations between
different NN pairs. It has long been used for surface diffusion with lateral
interactions~\cite{Reed1981}. It gives the NN correlation parameter $\eta$, defined by
$\avg{n_{\mathbf 0}n_{\mathbf L_j}}=\theta^2+\theta(1-\theta)\eta$. Its closed form was
derived in Ref.~\cite{PaperII} from the transfer matrix of the NN chain, with the
quasi-chemical condition~\cite{FowlerGuggenheim1939},
\begin{equation}
\eta\;=\;\frac{\bigl[1-4\theta(1-\theta)(1-e^{\beta\varepsilon})\bigr]^{1/2}-1}
{\bigl[1-4\theta(1-\theta)(1-e^{\beta\varepsilon})\bigr]^{1/2}+1} .
\label{eq:eta}
\end{equation}
A repulsion gives $\eta<0$. The probability that a NN site of an adparticle is occupied
is $q=\theta+(1-\theta)\eta$. The SSF is the static limit of the CF of Sec.~2,
$F(\K)=I(\K,0)=\sum_{\R}[\rho_{\R}(0)-\theta]e^{i\K\cdot\R}$. It is thus the CF of the
equilibrium distribution of separations between adparticles. Its zeroth moment
$F(\mathbf 0)=\sum_{\R}[\rho_{\R}(0)-\theta]$ is proportional to the compressibility of
the layer. Statics and dynamics are therefore described by the same CF. The RPA treats
every Fourier component of the occupation numbers as an independent Gaussian
mode~\cite{PaperII} and gives
\begin{equation}
F(\K)\;=\;\frac{1-\theta}{1-z\beta\varepsilon\,\theta(1-\theta)\,f_{\K}} .
\label{eq:ssf}
\end{equation}
Here $z$ is the number of NN sites of a site, $z=6$ on Pt(111), and
$\theta(1-\theta)$ is the variance of one occupation number. The interaction acts only
between NN sites, and its Fourier transform, normalized at $\K=0$, is $f_{\K}$. At
$\varepsilon=0$, $F=1-\theta$. A repulsion depresses $F$ at long wavelength, where
$f_{\K}>0$ and the layer becomes nearly incompressible. It raises $F$ where $f_{\K}<0$, most at
$\bar{\rm K}$, the wavevector of the $\sqrt3\times\sqrt3$ arrangement that a repulsive
triangular layer favours. At the nodes $K_0$ and $K_0'$, $f_{\K}=0$ and $F=1-\theta$ in
the RPA whatever the coupling. The RPA neglects the correlations between sites that are not
nearest neighbours. These correlations change $F$ at the nodes only slightly at low coverage. Away from
the nodes the $\K$ dependence of the SSF deforms the curve of the dephasing rate against
$\K$. Since only the coupling enters the SSF, this deformation depends on the coupling and not on the
friction of the substrate.

\section{Dynamics}

\subsection{A single adsorbate}

In the dilute limit the displacement of one adsorbate is a sum of independent jumps. Their
number in a time $t$ is Poisson distributed with mean $\Gamma_0t$, and each jump vector
$\mathbf L$ is drawn with probability $p_{\mathbf L}$. The displacement thus obeys a
compound Poisson distribution~\cite{TorresMiyares2025,TorresMiyares2026a}. Its CF is the
self ISF, $I/F=\exp\{-\Gamma_0t[1-f_{\K}]\}$. The ISF then decays as a single
exponential, and the dephasing rate is
\begin{equation}
\alpha(\K)\;=\;\Gamma_0\bigl[1-f_{\K}\bigr] .
\label{eq:ce}
\end{equation}
The same rate is given by the sum rule \eqref{eq:sumrule-gen} with
$\Gamma_{\rm eff}=\Gamma_0$ and $F=1$. Equation~\eqref{eq:ce} is the Chudley--Elliott
form~\cite{ChudleyElliott1961}. Its shape is fixed by the lattice through $f_{\K}$, and
its height is the total hop rate $\Gamma_0$. A single exponential means that the
collective friction vanishes in the dilute limit. The friction of the substrate is still
present, but only inside $\Gamma_0$.

For H and D on Pt(111) the three symmetry directions are equivalent, and the jumps along
each pair $\pm\mathbf L_j$ occur at the rate $\Gamma_0/3$. Equation~\eqref{eq:ce} then vanishes at the reciprocal-lattice
points and is largest at the zone boundary. On $\az{11\bar2}$, $\mathbf L_3$ is
orthogonal to $\K$, and
$\alpha=\tfrac23\Gamma_0[1-\cos(Ka\cos\tfrac{\pi}{6})]$. The CF analysis of a single
adsorbate reports $\tfrac23\Gamma_0$, the hop rate summed over the two directions seen on
this azimuth~\cite{TorresMiyares2025,TorresMiyares2026a}. The hop rates published from the
HSE data~\cite{Jardine2010} are the total hop rates $\Gamma_0$.

The single-adsorbate hop rate $\Gamma_0$ contains all the physics of the isolated adsorbate.
It is set first by the barrier along the jump, its height and its shape. Quantum
corrections modify it in two ways. The zero-point energies of the well and of the top
of the barrier change the effective barrier, and tunnelling through the barrier adds to
the thermal activation. The friction of the substrate, from the bath of substrate
excitations, also changes the rate. It acts on each adsorbate and is distinct from the
collective friction of Sec.~2, which acts on the density wave and whose bath is the
layer itself. Two regimes of the motion are separated by the crossover temperature
$T_c$. Above $T_c$ the adsorbate crosses the barrier mainly by thermal activation, with
quantum corrections. Below $T_c$ it passes mainly by tunnelling from the low states of
the well~\cite{Hanggi1990,Weiss1999}. For H on Pt(111), $T_c$ was estimated at about
$65$~K~\cite{Jardine2010}. Stochastic-wave-function calculations find, however, that
tunnelling carries most of the rate already below about $130$~K~\cite{TorresMiyares2025}. The memory equation needs none of these details. It needs
only that the jumps be incoherent and obey detailed balance. For the frictions estimated
for H on Pt(111) the tunnelling stays incoherent at every
temperature~\cite{WeissGrabert1985,Sanz2013,TorresMiyares2025}, and an adsorbate equilibrates
in its well within $7$--$14$~fs, long before it jumps again~\cite{TorresMiyares2025}.
Here $\Gamma_0$ is therefore treated as an unknown of the analysis, to be obtained from
the data.

\subsection{The adlayer}

At finite coverage the lattice gas needs a rate for each jump in each
configuration~\cite{PaperII}. This rate rests on three statements. First, the total
hop rate $\Gamma_0$ is shared equally among the $z$ NN jump vectors of Sec.~3. An
isolated adparticle thus jumps by one NN vector at the rate $\Gamma_0/z$. Second, let $\nu_{\R}$ be
the number of occupied NN sites of the departure site $\R$, other than the empty target.
Each of them adds the bond $-\varepsilon$ to the initial well. Third, the transition
state carries no bonds. The barrier counted from the initial well therefore changes by
$\varepsilon\nu_{\R}$, and the jump rate gains the Boltzmann factor
$e^{-\beta\varepsilon\nu_{\R}}$,
\begin{equation}
w_{\R}\;=\;\frac{\Gamma_0}{z}\,e^{-\beta\varepsilon\,\nu_{\R}} .
\label{eq:rate}
\end{equation}
For a repulsion an occupied neighbour raises the initial well and so lowers the
barrier. This jump rate obeys detailed balance with respect to \eqref{eq:H}, with either
coupling \eqref{eq:epsT}~\cite{PaperII}. It is the simplest such choice, since the
single-adsorbate theory does not fix how occupied neighbours modify a tunnelling jump
rate~\cite{WeissGrabert1985}. Another choice would change the hop rate in the layer but
neither the statics nor the sum rule. The rate also assumes that the friction of the
substrate does not depend on the environment, since it enters $\Gamma_0$ alone. $\Gamma_0$
multiplies every jump rate and only sets the time scale. Every rate of the signal is
therefore $\Gamma_0$ times a function of $\K$, $\theta$ and $\beta\varepsilon$ alone.

The effective hop rate $\Gamma_{\rm eff}$ of \eqref{eq:sumrule-gen} is given in the pair
approximation~\cite{PaperII},
\begin{equation}
\frac{\Gamma_{\rm eff}}{\Gamma_0}\;=\;(1-\theta)(1-\eta)\,\bigl[1-q+q\,e^{-\beta\varepsilon}\bigr]^{z-1}.
\label{eq:gc}
\end{equation}
The first two factors give the probability that the target is empty. The other $z-1$ NN
sites of the departure site are each occupied with the probability $q$, and the bracket is
the average of $e^{-\beta\varepsilon\nu_{\R}}$ over the binomial distribution of $\nu_{\R}$. A
repulsion raises $\Gamma_{\rm eff}$, since an adparticle with occupied neighbours leaves its
site faster. Each broken bond changes the barrier by $\varepsilon$, so that the mean change
of the barrier is $\varepsilon$ times the mean number of bonds broken. This is called bond
counting below. On Pt(111), to first order in $\beta\varepsilon$, the logarithm of
\eqref{eq:gc} grows by $-\beta\varepsilon\,\theta(6-\theta)$. For a constant coupling the
activation energy of $\Gamma_{\rm eff}$ thus falls by about $e_p\theta(6-\theta)$. Without
interaction $\Gamma_{\rm eff}=\Gamma_0(1-\theta)$ and $F=1-\theta$. The two factors cancel in
the sum rule, so that $\Omega=\Gamma_0[1-f_{\K}]$, the value of the dilute
limit~\cite{PaperII}. This ideal layer has no memory, $R=1$ and $\alpha=\Omega$. Any change
of the signal with coverage therefore comes from the interaction.

The closed forms give the statics and the local rate $\Omega$. The dephasing rate requires
the whole decay of the ISF, which no closed form gives. It is obtained from kinetic Monte
Carlo (KMC) simulations, which generate exact trajectories of the master equation with the
jump rates \eqref{eq:rate}. By \eqref{eq:alphaR} the dephasing rate is
$\lambda_{\min}\le\Omega/R$. The ratio $\lambda_{\min}/\Omega$ is the dynamical part of the
effect of the layer on the dephasing rate. It is fixed by the collective friction through
\eqref{eq:pole}, which is therefore written here explicitly. By \eqref{eq:laplace} the
memory function is given by the ISF,
\begin{equation}
\hat M(\K,s)\;=\;s+\Omega(\K)-\frac{F(\K)}{\hat I(\K,s)} .
\label{eq:Mhat}
\end{equation}
This expression gives the collective friction from any measured or simulated ISF. Its
lowest order is the single exponential of the second order of the hierarchy of
Ref.~\cite{PaperI},
\begin{equation}
M(\K,\tau)\;\simeq\;M(\K,0)\,e^{-\Omega_1(\K)\tau} ,
\label{eq:Mexp}
\end{equation}
with $M(\K,0)$ of \eqref{eq:M0}, the strength of the collective friction. The rate
$\Omega_1$ is fixed by the moments of the relaxation rates, as given in Appendix~A of
Ref.~\cite{PaperI},
$\Omega_1=\Omega+\sum_ka_k(\lambda_k-\Omega)^3/M(\K,0)$. It is the local rate shifted by the
asymmetry of the spread of the rates about $\Omega$. Like $\Omega$ and $M(\K,0)$, it is a
static average of the layer.
$1/\Omega_1$ is the memory time, the time over which the layer keeps
the memory of a jump. The integral of the memory function is then $M(\K,0)/\Omega_1$, and \eqref{eq:R}
gives the second-order value of $R$ in Refs.~\cite{PaperI,PaperII},
$[1-M(\K,0)/(\Omega\Omega_1)]^{-1}$. It contains the strength and the memory time only through
their ratio. The lowest rate separates them. It is the lowest root of the second-order
matrix of Ref.~\cite{PaperI}, the solution of \eqref{eq:pole} with this memory function,
\begin{equation}
\lambda_{\min}\;=\;\tfrac12\Bigl\{\Omega+\Omega_1-\bigl[(\Omega-\Omega_1)^2+4M(\K,0)\bigr]^{1/2}\Bigr\} .
\label{eq:lmin}
\end{equation}
Without collective friction $M(\K,0)=0$ and $\lambda_{\min}=\tau_c^{-1}=\Omega$. When the
memory time is short compared with the relaxation time of the density wave,
$\Omega_1\gg\Omega$, and $\lambda_{\min}\simeq\Omega-M(\K,0)/\Omega_1=\tau_c^{-1}$. In general
$\lambda_{\min}$ lies below $\tau_c^{-1}$, and both the strength and the memory time fix
the dephasing rate. KMC gives the memory function through \eqref{eq:Mhat}
(Appendix~\ref{app:kernel}).

The coupling is fitted to the QHAS coverage series by least squares. Each step of the
fit needs the dephasing rate, and thus $\lambda_{\min}/\Omega$, at every measured
temperature and coverage. A KMC simulation at each of these points and at each step
would be too costly. A closed form of $\lambda_{\min}/\Omega$ at the momentum transfer of
the series is therefore needed. Three properties of the memory fix its form. First,
$\lambda_{\min}=\Omega$ for the ideal layer. Second, the memory is of second order in the
coupling. At first order a weak coupling only shifts the rate of the single relaxation
mode of the ideal layer, and $\Omega$ already contains this shift. Other modes acquire
weight only at second order, since their weights are squares of first-order amplitudes.
The collective friction is therefore also of second order. Third, the memory vanishes for
an empty and for a full layer, where all sites are alike. The simplest form with these
properties is
\begin{equation}
\frac{\lambda_{\min}}{\Omega}\;=\;1-\kappa\,\theta(1-\theta)(\beta\varepsilon)^2 .
\label{eq:Rcf}
\end{equation}
It is an interpolation of the KMC results, not a theory of the memory. Its single
parameter $\kappa$ sets the strength of the memory and is fixed by the simulations.
At other momentum transfers KMC is used directly.

The aim of the analysis is the single-adsorbate hop rate $\Gamma_0$. A single-adsorbate
analysis obtains it by dividing the measured rate by the Chudley--Elliott factor
$1-f_{\K}$. At finite coverage this quotient is not $\Gamma_0$. The interaction factor $g$
quantifies the difference. It is the ratio of the dephasing rate of the layer to that of an isolated
adsorbate at the same $\K$ and $T$,
\begin{equation}
g(\K;\theta,\beta\varepsilon)\;=\;\frac{\alpha}{\Gamma_0\,[1-f_{\K}]}
\;=\;\Pi(\K;\theta,\beta\varepsilon)\,\frac{\lambda_{\min}}{\Omega} ,
\qquad
\Pi\;=\;\frac{\Gamma_{\rm eff}}{\Gamma_0\,F(\K)} .
\label{eq:g}
\end{equation}
The interaction factor $g$ depends on the coverage and the coupling but not on $\Gamma_0$. It
equals one without interaction, and a single-adsorbate analysis implicitly sets $g=1$.
Its separation from $\Gamma_0$ holds within the lattice gas with the jump rates
\eqref{eq:rate}, so that below the QHAS range $g$ is an estimate. The factorization
separates what the closed forms give from what needs the full dynamics. The static
factor $\Pi=\Omega/\{\Gamma_0[1-f_{\K}]\}$ is given by the closed forms. It corrects the
probability that an adparticle is ready to jump and the size of the density
fluctuations. The dynamical factor $\lambda_{\min}/\Omega$ is simulated by KMC. It gives the part
of the equilibrium flux $\Omega$ is undone by correlated back jumps, and it lies below
$1/R$. The friction of the substrate enters only through $\Gamma_0$, which cancels in $g$.
The collective friction enters only through $\lambda_{\min}/\Omega$, as a separate
factor.

The conversion from $\alpha$ to $\Gamma_0$ is simplest at the nodes $K_0$ and $K_0'$. There
$f_{\K}=0$, and in the RPA $F=1-\theta$ whatever the coupling. Let $g_0$ and $\Pi_0$ be
the values there. At the nodes $\Pi_0\simeq\Gamma_{\rm eff}/[\Gamma_0(1-\theta)]$ carries
no de Gennes narrowing, the slowing of the relaxation near the peak of the
SSF~\cite{deGennes1959}. A measurement at a node therefore gives the
hop rate nearly free of structure, $\Gamma_0=\alpha/g_0$.

To use the whole momentum dependence, the memory equation also fixes the form of the
curve of the dephasing rate against $\K$. For an isolated adsorbate this curve is the
Chudley--Elliott curve \eqref{eq:ce}, $\Gamma_0[1-f_{\K}]$. In the layer, $\Gamma_{\rm eff}$ does
not depend on $\K$, so that by \eqref{eq:g} the $\K$ dependence of $\Pi$ comes from
$1/F(\K)$ alone. The RPA \eqref{eq:ssf} gives
$1/F(\K)=[1-z\beta\varepsilon\theta(1-\theta)f_{\K}]/(1-\theta)$. With $\lambda_{\min}/\Omega$ replaced by
its value at the nodes, $g(\K)=g_0[1-z\beta\varepsilon\theta(1-\theta)f_{\K}]$, and
\begin{equation}
\alpha(\K)\;=\;\bar\alpha\,\bigl[1-f_{\K}\bigr]\bigl(1-\delta\,f_{\K}\bigr) ,
\qquad
\bar\alpha\;=\;\Gamma_0\,g_0 ,
\qquad
\delta\;=\;z\beta\varepsilon\,\theta(1-\theta) .
\label{eq:shape}
\end{equation}
The factor $z$ in the tilt is the number of NN sites, as in the SSF \eqref{eq:ssf}, and $\theta(1-\theta)$ is the variance of one occupation number. The rate $\bar\alpha$ is the height of the
curve, the dephasing rate at the nodes, and gives the hop rate,
$\Gamma_0=\bar\alpha/g_0$. The tilt $\delta$ quantifies the deformation of
the curve from the Chudley--Elliott curve. It is a new observable because the
Chudley--Elliott form has a height only. A single-adsorbate analysis thus extracts one
number from a curve, the hop rate. Equation~\eqref{eq:shape} extracts a second one, which
carries information on the layer and none on $\Gamma_0$. It keeps the zeros of $1-f_{\K}$,
fixed by the lattice, and changes only the weight of long against short wavelengths. For
the ideal layer $\delta=0$, and \eqref{eq:shape} reduces to \eqref{eq:ce}. A repulsion
gives $\delta<0$. The dephasing rate then exceeds the Chudley--Elliott curve at long
wavelength, where the nearly incompressible layer relaxes a density wave faster than an
isolated adsorbate. It falls below it at the zone boundary, the de Gennes narrowing.
Equation~\eqref{eq:shape} is linear in $\bar\alpha$ and $\bar\alpha\delta$. A weighted
linear fit to a measured curve gives both with their errors, without a model of the
layer. The published analyses fitted the Chudley--Elliott form, which sets the tilt to
zero, so that any published curve can be reanalysed for it. At strong coupling the RPA overestimates the
tilt. KMC gives it exactly, with the $\K$ dependence of the SSF and of the collective
friction (Appendix~\ref{app:kmctilt}). The friction of the substrate enters only
$\bar\alpha$. Within the lattice gas the tilt thus depends on the coupling alone. A tilt
that the coupling cannot produce signals physics outside the model.

The CF method gives the diffusion coefficients from the long-wavelength limit of the
CF~\cite{TorresMiyares2026a}. For an isolated adsorbate the CF of the displacement is
$\exp(-\tfrac12K^2\avg{L_\parallel^2})$ at small $K$, with $L_\parallel$ the projection of
the displacement on the direction of observation, the direction of $\K$. As in
Ref.~\cite{TorresMiyares2026a}, Einstein's law $\avg{L_\parallel^2}=D_Kt$ defines the
single-adsorbate coefficient along this direction. With \eqref{eq:ce},
$D_K=(\Gamma_0/K^2)\sum_{\mathbf L}p_{\mathbf L}(\K\cdot\mathbf L)^2$. Only the jumps not
orthogonal to $\K$ contribute, each through its projection on the direction of
observation. The coefficient therefore depends on the azimuth through the directions that contribute
and their projections, that is, through the scalar products $\K\cdot\mathbf L$.
On $\az{11\bar2}$, the azimuth of the HSE data, the pairs $\pm\mathbf L_1$ and
$\pm\mathbf L_2$ contribute with the projection $a\cos\psi_1$, $\psi_1=\pi/6$, and
$\pm\mathbf L_3$ drops out. With the rate $\Gamma_0/3$ of each pair,
\begin{equation}
D_K\;=\;2\,\frac{\Gamma_0}{3}\,a^2\cos^2\psi_1 ,
\label{eq:DK}
\end{equation}
the expression of Ref.~\cite{TorresMiyares2026a}. On $\az{1\bar10}$ the three pairs
contribute, each with its own projection, and $D_K$ is given by the same sum with these
projections. Ref.~\cite{Jardine2010} counts the three symmetry directions in the hop rate, which
is the total hop rate $\Gamma_0$, and uses the isotropic coefficient $a^2\Gamma_0/4$ of
the motion in the plane. This coefficient contains no projection on the direction of
observation and cannot coincide with $D_K$. At finite $K$ the two azimuths give
different curves, as Fig.~\ref{fig:directions}(b) shows. The same route applies to the coherent CF \eqref{eq:cf}. For small $K$,
$I(\K,t)=F(\mathbf 0)-\tfrac12K^2M_2(t)$, with
$M_2(t)=\sum_{\R}R_\parallel^2[\rho_{\R}(t)-\theta]$ the second moment of the separation
distribution and $R_\parallel$ the component of $\R$ along $\K$. The zeroth moment
$F(\mathbf 0)$ does not change in time, since the adparticles are conserved. In the
diffusive regime $I\simeq F(\K)\,e^{-\frac12D_cK^2t}$, with $D_c$ the collective diffusion
coefficient. Comparing the terms in $K^2$ gives Einstein's law for the layer,
\begin{equation}
M_2(t)-M_2(0)\;=\;F(\mathbf 0)\,D_c\,t .
\label{eq:einstein}
\end{equation}
The growth of the second moment equals the mean square displacement of the centre of
mass, $\avg{(\sum_iL_{\parallel,i})^2}/N$, with the sum over the $N$ adparticles. $D_c$
is thus the centre-of-mass coefficient divided by $F(\mathbf 0)$. At short times the
jumps are uncorrelated, and the slope of $M_2$ gives the long-wavelength limit of the
local rate, $\Omega\to(\Gamma_{\rm eff}/\Gamma_0)\,D_KK^2/2F(\mathbf 0)$~\cite{PaperII}.
At long times correlated back jumps can reduce the slope. At long wavelength the
diffusive mode carries the whole weight, so that the ratio of the two slopes is
$\lambda_{\min}/\Omega=1/R(\mathbf 0)$, the dynamical factor at long wavelength. Hence
\begin{equation}
D_c\;=\;D_K\,\frac{\Pi(\mathbf 0)}{R(\mathbf 0)} ,
\qquad
\Pi(\mathbf 0)\;=\;\frac{\Gamma_{\rm eff}}{\Gamma_0F(\mathbf 0)} .
\label{eq:Dc}
\end{equation}
The collective coefficient is the single-adsorbate coefficient times the interaction
factor $g$ at long wavelength. $D_K$ contains the friction of the substrate. $\Pi(\mathbf 0)$
contains the static part, with the thermodynamic factor $1/F(\mathbf 0)$. $R(\mathbf 0)$
contains the collective friction. Without interaction $\Gamma_{\rm eff}=\Gamma_0(1-\theta)$
and $F(\mathbf 0)=1-\theta$, so that $D_c=D_K$ at every coverage~\cite{PaperII,Kutner1981}.
KMC gives $R(\mathbf 0)$ from the centre-of-mass displacement
(Appendix~\ref{app:diff}).

\section{Analysis of the measured dephasing rates}

Two kinds of data are analysed. The first is the QHAS coverage series, measured at a
single momentum transfer at five coverages~\cite{Graham1999}. It fixes the lateral
coupling. The second consists of eleven momentum-resolved curves, four from HSE and
seven from QHAS~\cite{Jardine2010,Graham1999}. They give the hop rates and test the
coupling. The first kind gives a rate at a single momentum transfer and no tilt. The
second gives the whole curve, its height and its tilt. The curves and their error bars were digitized from the published figures,
as described in Appendix~\ref{app:digit}. Throughout the analysis the friction of the
substrate is contained in the single-adsorbate hop rate $\Gamma_0$, and the collective
friction in the dynamical factor $\lambda_{\min}/\Omega$.

\subsection{The adlayer in the simulations}

This subsection describes the KMC simulations and the properties of the adlayer obtained
from them. It contains no measured data. All quantities in it are simulated, for a range
of couplings, and the analysis of the measured rates relies on them.

The simulations are set up in this way. The adparticles occupy a triangular lattice of
$60\times60$ sites with periodic boundaries. At the coverage $\theta$ there are
$3600\,\theta$ adparticles, placed at random at the start. An adparticle jumps to an empty
NN site with the rate \eqref{eq:rate}. At each step one jump is chosen with a probability
proportional to its rate. The time then advances by a random interval drawn from an
exponential distribution, whose mean is the inverse of the total rate of all
jumps~\cite{Bortz1975}. Since $\Gamma_0$ multiplies every jump rate, it is set to one, and
all times are in units of $1/\Gamma_0$. A run starts with an equilibration over $50$ time
units. The density waves $\delta n_{\K}$ of \eqref{eq:dw} are then recorded every $0.01$
units over $5000$ units. They are recorded at all momentum transfers allowed by the
periodic lattice, along the three equivalent directions of each azimuth. Their correlation
gives the ISF, averaged over the time origins, the three directions and independent runs.
Each run thus covers thousands of jumps of each adparticle. The statistical errors
quoted below are standard errors over the runs.

Every quantity of Secs.~2--4 is obtained from these data. $\Gamma_{\rm eff}$ is the number
of jumps per unit time and adparticle. The SSF is the ISF at $t=0$. The local rate
$\Omega$ is given by $\Gamma_{\rm eff}$ and $F$ through the sum rule \eqref{eq:sumrule}.
The correlation time $\tau_c$ is the area beneath $I/F$, and $R=\Omega\tau_c$. The
dephasing rate is obtained as in the HSE experiment, from a single exponential fitted to
$I/F$ between $0.3$ and $0.05$, a window in the diffusive regime. The tilt of a simulated
curve is obtained by fitting \eqref{eq:shape} to the simulated dephasing rates against
$K$. Sixteen runs are averaged for each coverage and coupling, and the statistical error
of a tilt is then about $0.02$. The memory function is obtained from \eqref{eq:Mhat} at the
node $K_0$ of $\az{11\bar2}$, with six runs for each case. The Laplace transform
$\hat I(\K,s)$ is computed for $0\le s\le3\Omega$ by integrating the ISF exactly between
the recorded times. The tail beyond $I/F=0.01$ is continued with the lowest rate, fitted
for this purpose between $0.2$ and $0.02$, deeper in the diffusive regime. The single
exponential \eqref{eq:Mexp} is then fitted to $\hat M$, with $M(\K,0)$ and $\Omega_1$ as
parameters. The long-wavelength factor $1/R(\mathbf 0)$ of \eqref{eq:Dc} comes from eight
separate runs of $3\times10^4$ units, or four shorter runs at $0.4$~ML and
$\beta\varepsilon=-0.88$, that record the displacement of the centre of mass.
It is the long-time slope of the mean square displacement of the centre of mass divided
by its short-time slope. The values of $\lambda_{\min}/\Omega$ that fix $\kappa$ come
from one run of $2\times10^4$ units for each coverage and coupling.

The simulations first test the identification \eqref{eq:alphaR}. At $0.1$~ML the slowest
mode carries the weight $a_{\min}=0.83$--$0.85$, the intercept at $t=0$ of the exponential
fitted in the diffusive regime. The rate fitted in the window of the experiments is
$1.03$--$1.05$ times $\lambda_{\min}$ at $0.1$ and $0.4$~ML. The difference is smaller than
the experimental errors. The window rate is thus $\lambda_{\min}$, as stated in
Refs.~\cite{PaperI,PaperII}.

The memory function is represented by the single exponential \eqref{eq:Mexp}. The largest relative
deviation between the two is $1.5$\% over $0\le s\le3\Omega$. Table~\ref{tab:kernel} of the Appendix gives the fitted values. The
memory time $1/\Omega_1$ is $0.3$--$0.5$ times the relaxation time $1/\lambda_{\min}$ of the
density wave. The integral of the memory function gives
$\hat M(\K,0)/\Omega=1-1/R$, the fraction of $\Omega$ that it removes. This fraction is
$0.085$ and $0.15$ at $0.1$~ML and $0.16$ and $0.27$ at $0.4$~ML, for
$\beta\varepsilon=-0.60$ and $-0.88$. The inverse correlation time $\tau_c^{-1}=\Omega/R$
is then $1.10$--$1.13$ times $\lambda_{\min}$ at $0.1$~ML and up to $1.35$ times at
$0.4$~ML. The collective friction thus lowers the dephasing rate more than $R$ alone
indicates. With the fitted strength and memory time, \eqref{eq:lmin} gives $1.005$--$1.03$
times the simulated $\lambda_{\min}$ at $0.1$~ML and $1.05$--$1.14$ times at $0.4$~ML. The
collective friction therefore has memory, and both its strength and its memory time
matter.

At long wavelength, in contrast, $1/R(\mathbf 0)$ lies between $0.99$ and $1.01$ for
$\beta\varepsilon$ between $-0.29$ and $+0.02$ and $\theta=0.1$--$0.6$, with a statistical
error of $0.008$. At $\beta\varepsilon=-0.88$ it is $0.98\pm0.01$ at $0.2$~ML and
$1.00\pm0.01$ at $0.4$~ML. The collective friction thus nearly vanishes at long
wavelength. The back jumps are local, and their effect averages out over a long density
wave. The enhancement of collective diffusion by a repulsion is therefore static. It
comes from the low compressibility, through $1/F(\mathbf 0)$.

The closed form \eqref{eq:Rcf} is fitted to $46$ simulated values of $\lambda_{\min}/\Omega$
at $0.45$~\AA$^{-1}$ on $\az{1\bar10}$, next to the $0.44$~\AA$^{-1}$ of the QHAS coverage
series. They cover $\theta=0.1$--$0.66$ and $\beta\varepsilon$ from $-1.40$ to $0.07$. The fit
gives $\kappa=2.19\pm0.13$. The simulated values scatter about \eqref{eq:Rcf} by $0.05$ in
$\lambda_{\min}/\Omega$, mainly from the statistical noise of the window fits. This
corresponds to an uncertainty of about $5$\% in the calculated widths. In the range of the
coverage series, $|\beta\varepsilon|\le0.35$, \eqref{eq:Rcf} gives
$\lambda_{\min}/\Omega\ge0.93$.

At the nodes the simulated $F/(1-\theta)$ lies between $0.96$ and $1.00$ at $0.1$~ML and
falls to $0.84$ at $0.4$~ML and $\beta\varepsilon=-0.88$. The RPA gives exactly one there.
The difference comes from correlations between sites that are not nearest neighbours,
which the RPA neglects. The simulated tilt is small at low coverage. At $\theta=0.1$ it
lies between $0$ and $-0.30$ for $\beta\varepsilon$ from $0$ to $-1.4$
(Appendix~\ref{app:kmctilt}).

\subsection{The lateral coupling}

The lateral coupling is fixed by the QHAS coverage series of Ref.~\cite{Graham1999}. The
measured quantity is the quasielastic width $\Gamma_{\rm QHAS}=2\hbar\alpha$ at
$0.44$~\AA$^{-1}$ on $\az{1\bar10}$, between $130$ and $250$~K. Graham et al.\ fitted the
measured widths at each of five coverages, from $0.1$ to $0.66$~ML, to an Arrhenius law
$\nu_0e^{-E_d/k_BT}$. Here $E_d$ is the activation energy and $\nu_0$ the prefactor, given
in frequency units as in Ref.~\cite{Graham1999}. These published values of $E_d$ and
$\nu_0$ are not measured directly. They summarize the measured widths at each coverage
by two numbers, with error bars. Both fall with coverage, whereas the widths at
different coverages nearly coincide.

The lattice gas gives the width at every temperature and coverage. It is called here the
calculated width, $2\hbar\alpha$ with $\alpha=\Gamma_0\,g\,[1-f_{\K}]$ and
$g=\Pi\,\lambda_{\min}/\Omega$ from \eqref{eq:g}. The static factor $\Pi$ is given by
\eqref{eq:gc} and \eqref{eq:ssf}, and the dynamical factor $\lambda_{\min}/\Omega$ by
\eqref{eq:Rcf}. Between $130$ and $250$~K the single-adsorbate hop rate $\Gamma_0$ is
represented by an Arrhenius law, with its own activation energy and prefactor. These two
parameters absorb the quantum corrections of the hop rate. The calculated widths at each
coverage are then fitted to an Arrhenius law over the same range, as Graham et al.\ did
with the measured widths. This gives the calculated $E_d$ and $\nu_0$. Their dependence
on the coverage comes from $g$ alone, and their values at low coverage from $\Gamma_0$.

The fit has four parameters, the two Arrhenius parameters of $\Gamma_0$ and the two
parameters of the coupling, $e_p$ and $T^*$. A constant coupling has $e_p$ only. The constant coupling is the usual assumption, used in
Ref.~\cite{Graham1999}. It is the alternative against which the temperature-dependent
coupling is tested, here and with the tilts of the momentum-resolved curves. Outside these tests it is not used. The
parameters are fitted to the five published activation energies and five prefactors by
minimizing $\chi^2$, the sum of the squared deviations in units of the error bars,
given in \eqref{eq:chi2} of Appendix~\ref{app:chi2}. The number of degrees of freedom
is the number of fitted values minus the number of parameters. For a correct model
$\chi^2$ is of the order of this number. Let $P$ be the probability that a correct model
gives a $\chi^2$ larger than the one found. A small $P$ means that the deviations are too
large to come from the errors alone. A fit is rejected when $P$ is below $0.05$.

The two couplings of \eqref{eq:epsT} act differently on the Arrhenius parameters. By
bond counting (Sec.~4.2), each bond broken in a jump changes the barrier by
$\varepsilon$. With a constant coupling, $-\beta\varepsilon=e_p/k_BT$ is a pure $1/T$
term. Each broken bond then lowers $E_d$ by $e_p$ and leaves $\nu_0$ unchanged. With the
temperature-dependent coupling, $-\beta\varepsilon=e_p/k_BT-s_p/k_B$. The first term
lowers $E_d$, and the second, independent of temperature, lowers $\ln\nu_0$. Both are
proportional to the same mean number of bonds and obey the compensation rule of Sec.~3,
$\Delta E_d=k_BT^*\Delta\ln\nu_0$. The published values obey such a rule before any
model is fitted. Between $0.1$ and $0.6$~ML, $E_d$ falls by $28$~meV and $\ln\nu_0$ by
$1.51$. Their ratio corresponds to the temperature $215$~K.

\begin{figure}[!t]
\centering
\includegraphics[width=0.99\textwidth]{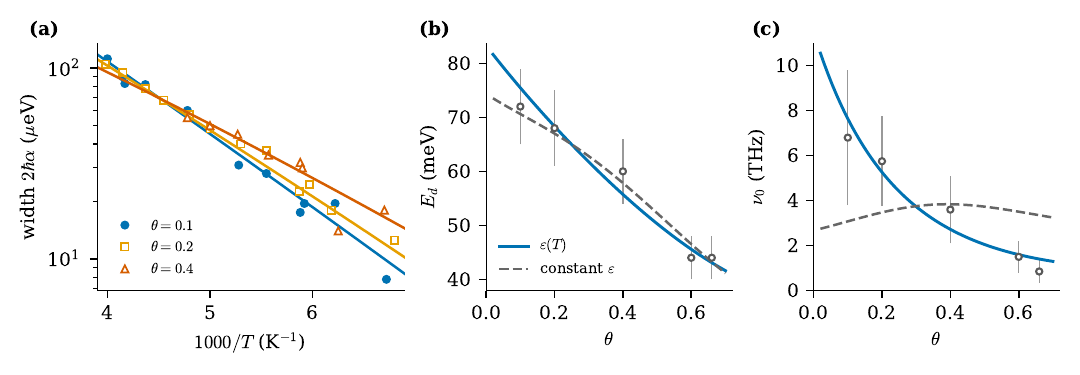}
\caption{The lattice gas against the QHAS coverage series of Ref.~\cite{Graham1999},
at $0.44$~\AA$^{-1}$ on $\az{1\bar10}$. (a) Measured widths
$\Gamma_{\rm QHAS}=2\hbar\alpha$ at $\theta=0.1$, $0.2$ and $0.4$ (symbols), digitized
without error bars, and calculated widths for the temperature-dependent coupling (lines).
(b) Activation energy $E_d$ and (c) prefactor $\nu_0$ of Arrhenius fits between $130$ and
$250$~K. Circles are the published values of Ref.~\cite{Graham1999}, obtained from the
measured widths, with their error bars. Lines are the same fits of the calculated widths,
for $\varepsilon(T)$ with $e_p=9.5$~meV and $T^*=222$~K (solid) and for the constant
coupling $\varepsilon=-8.6$~meV (dashed). The calculated widths are obtained from
\eqref{eq:sumrule}, \eqref{eq:ssf}, \eqref{eq:gc} and \eqref{eq:Rcf}. Fitted are the two
Arrhenius parameters of the single-adsorbate hop rate and the parameters of the coupling,
to (b) and (c). With the coupling fixed, the two Arrhenius parameters are then refitted
to (a). The calculated widths use the closed form \eqref{eq:Rcf}, whose parameter $\kappa$
is fixed by the simulations.}
\label{fig:coverage}
\end{figure}

Figure~\ref{fig:coverage} shows that only the temperature-dependent coupling reproduces
the coverage series. Panel (a) displays the measured and calculated widths, and panels
(b) and (c) $E_d$ and $\nu_0$ against coverage. The temperature-dependent coupling gives
$e_p=9.5\pm1.9$~meV, $T^*=222\pm73$~K and $s_p=0.50\pm0.14\,k_B$, the error of $s_p$
including the correlation of $e_p$ and $T^*$. Its $\chi^2$ is $2.3$ for $6$ degrees of
freedom, with $P=0.89$. The constant coupling, $\varepsilon=-8.6$~meV, gives $\chi^2=14.7$
for $7$ degrees of freedom, with $P=0.04$, and is disfavoured. In panel (b) both couplings
reproduce the fall of $E_d$. About four fifths of it is bond counting. The rest comes
from the fall of $F$ at $0.44$~\AA$^{-1}$, where $f_{\K}>0$. In panel (c) only the
temperature-dependent coupling reproduces the fall of $\nu_0$. A constant coupling leaves
the prefactor near $3$--$4$~THz.

The measured widths themselves are not used to fit the coupling. They test it. With the
coupling fixed, the two Arrhenius parameters of $\Gamma_0$ are refitted to the $31$
measured widths of panel (a). These widths carry no published error bars, so relative
errors are assumed, rising from $15$\% at $250$~K to $45$\% at $130$~K. The
temperature-dependent coupling gives $\chi^2=5.6$ for $29$ degrees of freedom, and the
constant coupling $50.5$, with $P=0.008$. The assumed errors are generous, so that the
small value $5.6$ only shows that they are too large. The comparison of the two values
is the meaningful result. In panel (a) the temperature-dependent coupling reproduces the
nearly coinciding widths with a single hop rate, and the constant coupling misses them.

The parameters of $\varepsilon(T)$ are robust, but the evidence against the constant
coupling from the published $E_d$ and $\nu_0$ alone is marginal. Leaving out one coverage
at a time gives $e_p$ between $8.9$ and $10.6$~meV and $T^*$ between $186$ and $268$~K, with
$P\ge0.70$ for $\varepsilon(T)$. For the constant coupling $P$ then lies between $0.01$ and
$0.13$. If the published errors were $1.5$ times larger, $\chi^2$ would fall by the factor
$2.25$, and the constant coupling would give $P=0.48$. The case for $\varepsilon(T)$
therefore rests on three indications together. They are the published $E_d$ and $\nu_0$,
the measured widths and the tilts of the momentum-resolved curves. The size of the pair
entropy adds a fourth, since a softening of the well by an occupied neighbour can explain
it (Sec.~3).

The fitted pair energy is close to the repulsion inferred before from
$E_d$~\cite{Graham1999} and to the $8.7$~meV of H/Ni(111)~\cite{Cao1997}. In the QHAS
range the reduced coupling stays small, from $-0.35$ at $130$~K to $+0.06$ at $250$~K.
The layer is therefore nearly ideal there, and the widths coincide near $T^*$, as Graham
et al.\ noticed~\cite{Graham1999}. A compensation can also appear as an artefact of
Arrhenius fits. The fitted slope and intercept of a straight line in $1/T$ have
correlated errors. Lines fitted to noisy data therefore pivot about the centre of the
data in $1/T$, the mean of $1/T$ over the measured temperatures. The inverse of this mean
is the harmonic mean of the temperatures, $182$~K here. A compensation produced by the
errors alone would lie there. The compensation temperature of the data, $215$~K, and the
fitted $T^*$ lie above it. The difference, $33$ to $40$~K, is smaller than the
uncertainty of $T^*$, $73$~K, so that this argument alone does not exclude an artefact.

\subsection{The interaction factor below the QHAS range}

The HSE Arrhenius plot at $0.1$~ML extends down to $80$~K, below the QHAS range of
$130$ to $250$~K where the coupling was fitted. To convert these measured rates into
single-adsorbate hop rates, the interaction factor $g$ is needed down to $80$~K. This
subsection gives it. The coupling $\varepsilon(T)$ of Sec.~5.2 is extended below
$130$~K with \eqref{eq:epsT}. At $80$~K it gives $\beta\varepsilon=-0.88$. The interaction
factor is then simulated by KMC at each temperature, at the node $K_0$ of $\az{11\bar2}$,
where the HSE rates were measured. Below $130$~K tunnelling dominates the
jumps~\cite{TorresMiyares2025}, and the jump rates \eqref{eq:rate} are an assumption for
tunnelling jumps. Below the QHAS range the interaction factor is therefore an
extrapolation of the coupling and of the jump rates.

At the node the interaction factor is $g_0$, the ratio of the dephasing rate of the layer
to that of an isolated adsorbate with the same hop rate $\Gamma_0$. A value $g_0=1.17$, for
instance, means that the layer raises the dephasing rate by the factor $1.17$. A
single-adsorbate analysis, which sets $g_0=1$, then overestimates $\Gamma_0$ by the same
factor.

\begin{figure}[!t]
\centering
\includegraphics[width=0.99\textwidth]{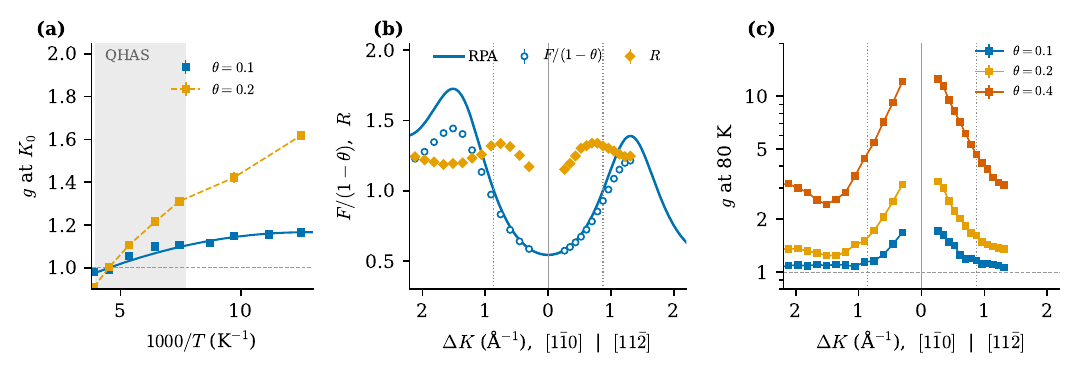}
\caption{Interaction factor below the QHAS range. (a) Value $g_0$ of the interaction factor
at the node $K_0$ of $\az{11\bar2}$ against $1000/T$, at $\theta=0.1$ (blue) and $0.2$
(orange). (b) SSF divided by $1-\theta$ (circles and line) and $R=\Omega\tau_c$ (diamonds)
against $K$ at $80$~K and $\theta=0.2$, on $\az{1\bar10}$ (left) and $\az{11\bar2}$ (right).
(c) Interaction factor $g(\K)$ against $K$ at $80$~K for $\theta=0.1$, $0.2$ and $0.4$.
All symbols are simulated by KMC, with the reduced coupling given at each temperature by
the coupling $\varepsilon(T)$ fitted in Sec.~5.2. In (a) the line is a quadratic fit in
$\beta\varepsilon$ to the simulated points at $\theta=0.1$, and the shaded band is the QHAS
range. In (b) the circles are the simulated ISF at $t=0$. The diamonds combine $\Omega$,
from the sum rule \eqref{eq:sumrule} with the simulated $\Gamma_{\rm eff}$ and $F$, and
$\tau_c$, the area beneath the simulated $I/F$. The line is the SSF calculated with the RPA
\eqref{eq:ssf} for $\theta=0.2$ and $\beta\varepsilon=-0.88$. In (c) the lines join the
simulated points, and dotted lines mark the nodes $K_0'$ and $K_0$. Error bars are standard
errors over $16$ independent runs and are smaller than the symbols where not visible. No
measured data enter this figure.}
\label{fig:memory}
\end{figure}

Figure~\ref{fig:memory} shows how the layer changes the dephasing rate below the QHAS
range. Panel (a) displays $g_0$ at the node against $1000/T$, at $0.1$ and $0.2$~ML.
Panel (b) displays two properties of the layer that enter $g$, the SSF and the memory
measure $R$, against $K$ at $80$~K and $0.2$~ML on both azimuths. Panel (c) displays the
interaction factor $g(\K)$ against $K$ at $80$~K, for $0.1$, $0.2$ and $0.4$~ML. All
symbols are simulated. The only calculated curve is the RPA in panel (b). No measured data enter the figure. These simulated values are used to convert the
measured rates into hop rates.

Panel (a) shows that $g_0$ stays close to one at $0.1$~ML and grows with the coverage.
In the QHAS range the simulated $g_0$ lies between $0.98$ and $1.11$, since the reduced coupling is
small there. Below it, $g_0$ rises at $0.1$~ML to $1.17$ at $80$~K,
and at $0.2$~ML to $1.62$ at $80$~K. Two factors of \eqref{eq:g} compete, both simulated. At $80$~K the
static factor $\Pi_0$ is $1.51$ at
$0.1$~ML and $2.52$ at $0.2$~ML. It exceeds one because
an adparticle with occupied neighbours leaves its site faster. The dynamical factor
$\lambda_{\min}/\Omega$ is $0.77$ and $0.64$. It lies below one because of the collective
friction. Their product is $g_0$. At $0.1$~ML and $80$~K the memory time $1/\Omega_1$ is
about $0.3$ times the relaxation time $1/\lambda_{\min}$ of the density wave
(Table~\ref{tab:kernel}).

Panel (b) shows the origin of the momentum dependence of $g$. The repulsion depresses the
simulated SSF at long wavelength, where the layer is nearly incompressible, and raises it near the
zone boundary, where the layer favours the $\sqrt3\times\sqrt3$ arrangement. The calculated RPA
locates the maxima but overestimates them. The simulated $R$ lies between $1.15$ and $1.4$. It is the
memory measure of Refs.~\cite{PaperI,PaperII}, and $1/R$ is an upper bound of
$\lambda_{\min}/\Omega$.

Panel (c) shows that the simulated $g(\K)$ rises towards long wavelength at every coverage. At the
smallest momentum transfer it reaches $1.7$ at $0.1$~ML, $3.3$ at $0.2$~ML and $12$ at
$0.4$~ML. There the low compressibility raises the collective diffusion. This rise is
static, since the collective friction nearly vanishes at long wavelength (Sec.~5.1). At
$0.4$~ML, $g(\K)$ dips at $\bar{\rm K}$ on $\az{1\bar10}$. This is the de Gennes narrowing
at the peak of the SSF. At the node and $0.4$~ML, $g_0\approx4.7$ at $80$~K. A
single-adsorbate analysis would overestimate $\Gamma_0$ by this factor.

These values rest on the jump rates \eqref{eq:rate}. They assume that the occupied
neighbours of the departure site lower the barrier, whereas the neighbours of the target
do not matter. Detailed balance fixes only the ratio of a jump and its reverse. For
incoherent tunnelling the rate depends on the energy difference between the initial and
the final site~\cite{WeissGrabert1985,Weiss1999}. This suggests a symmetric rule,
$w=(\Gamma_0/z)\,e^{-\beta\varepsilon(\nu_{\R}-\nu_{\rm T})/2}$, with $\nu_{\rm T}$ the number
of occupied NN sites of the target other than the departure site. It obeys detailed
balance with the same energy \eqref{eq:H} and gives the same statics. In the QHAS range
the data decide between the two rules. With the symmetric rule the calculated widths do
not reproduce the coverage series. The temperature-dependent coupling then gives
$\chi^2=15.2$ for $6$ degrees of freedom, $P=0.02$, with a pair energy at $30$~meV, the upper
bound allowed in the fit, and the constant coupling $\chi^2=35.5$ for $7$. The fall of $E_d$ by bond
counting requires the rule \eqref{eq:rate}. Below $130$~K, where tunnelling dominates, no
theory decides between them, and the symmetric rule was simulated as well.

\begin{table}[!t]
\centering
\caption{Static factor $\Pi_0$, dynamical factor $\lambda_{\min}/\Omega$, interaction factor
$g_0$ at the node $K_0$ of $\az{11\bar2}$, and tilt $\delta$ of the curve on $\az{11\bar2}$, for the jump rates
\eqref{eq:rate} and for the symmetric rule. All values are simulated by KMC with $16$
runs, except those of \eqref{eq:rate} at $0.4$~ML, with $3$ runs. The temperatures are
those of the fitted coupling. The statistical errors are about $0.02$ in $g_0$ and $0.02$
in $\delta$.}
\label{tab:rule}
\vspace{4pt}
\small
\begin{tabular}{c c c c c c c c c c}
\hline
 & & & \multicolumn{4}{c}{Rule \eqref{eq:rate}} & \multicolumn{3}{c}{Symmetric rule}\\
$\theta$ & $\beta\varepsilon$ & $T$ (K) & $\Pi_0$ & $\lambda_{\min}/\Omega$ & $g_0$ & $\delta$ & $\Pi_0$ & $\lambda_{\min}/\Omega$ & $g_0$\\
\hline
$0.1$ & $-0.32$ & $135$ & $1.19$ & $0.93$ & $1.11$ & $-0.13$ & $1.02$ & $1.01$ & $1.03$\\
$0.1$ & $-0.57$ & $103$ & $1.33$ & $0.86$ & $1.15$ & $-0.18$ & $1.03$ & $0.97$ & $1.00$\\
$0.1$ & $-0.88$ & $80$ & $1.51$ & $0.77$ & $1.17$ & $-0.27$ & $1.04$ & $0.94$ & $0.97$\\
$0.2$ & $-0.32$ & $135$ & $1.43$ & $0.92$ & $1.31$ & $-0.26$ & $1.04$ & $1.00$ & $1.04$\\
$0.2$ & $-0.57$ & $103$ & $1.86$ & $0.76$ & $1.42$ & $-0.41$ & $1.08$ & $0.97$ & $1.04$\\
$0.2$ & $-0.88$ & $80$ & $2.52$ & $0.64$ & $1.62$ & $-0.59$ & $1.11$ & $0.93$ & $1.02$\\
$0.4$ & $-0.32$ & $135$ & $2.10$ & $0.91$ & $1.91$ & $-0.47$ & $1.08$ & $0.99$ & $1.07$\\
$0.4$ & $-0.88$ & $80$ & $8.07$ & $0.58$ & $4.67$ & $-1.14$ & $1.22$ & $0.96$ & $1.18$\\
\hline
\end{tabular}
\end{table}

Table~\ref{tab:rule} compares the two rules. The simulated SSF is the same for both, as it
must be. With the symmetric rule the neighbours of the departure site and of the target
nearly cancel. $\Pi_0$ then stays between $1.02$ and $1.22$, the collective friction is
weaker, and $g_0$ lies between $0.97$ and $1.18$. The two rules thus bracket $g_0$ below the
QHAS range. At $0.1$~ML and $80$~K, $g_0$ lies between $0.97$ and $1.17$. The extracted hop
rates there carry a systematic uncertainty of a factor of up to $1.2$. At $0.2$~ML the
bracket is $1.02$ to $1.62$, and at $0.4$~ML $1.18$ to $4.67$. The tilt changes by at most $0.08$ with
the rule. With the symmetric rule it is $-0.12$, $-0.24$ and $-0.35$ at $0.1$~ML and
$-0.31$, $-0.49$ and $-0.67$ at $0.2$~ML, for the three couplings of the table. It comes
mainly from the SSF.

\subsection{The momentum-resolved rates}

This subsection analyses the eleven momentum-resolved curves, four from HSE and seven
from QHAS. Each curve is reduced to two numbers, the node rate $\bar\alpha$ and the tilt
$\delta$ of \eqref{eq:shape}. The node rate gives the hop rate. The tilt is a test of the
coupling fitted in Sec.~5.2. The two numbers differ much in how well the data fix them.

\begin{figure}[!t]
\centering
\includegraphics[width=0.99\textwidth]{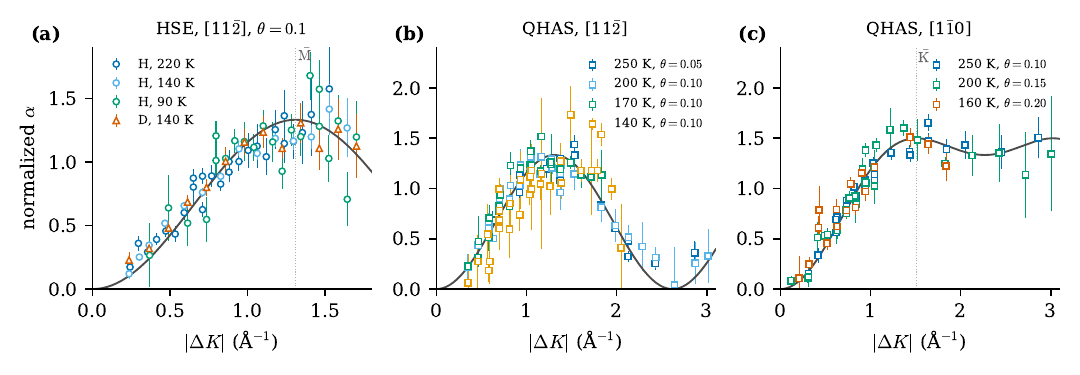}
\caption{Momentum-resolved dephasing rates of H and D on Pt(111), divided by the
height of the Chudley--Elliott form fitted to each curve. (a) HSE on
$\az{11\bar2}$ at $0.1$~ML~\cite{Jardine2010}. (b) QHAS on
$\az{11\bar2}$~\cite{Graham1999}. (c) QHAS on $\az{1\bar10}$~\cite{Graham1999}. The
temperatures and coverages are given in the legends and in Table~\ref{tab:kfit}. Lines
show the ideal-layer shape $1-f_{\K}$, calculated from the lattice geometry. Dotted
lines mark the zone-boundary points $\bar{\rm M}$ and $\bar{\rm K}$. Symbols are
measured and digitized. Error bars are the published bars, digitized and divided by
$\bar\alpha$. The only fitted quantity is $\bar\alpha$ of each curve. No simulation enters
this figure.}
\label{fig:kcurves}
\end{figure}

Figure~\ref{fig:kcurves} shows that every measured curve has nearly the shape of the
ideal layer. Each panel displays the measured rates divided by the height $\bar\alpha$ of
the Chudley--Elliott form fitted to the curve, together with the calculated shape
$1-f_{\K}$. Panel (a) displays the HSE curves, and panels (b) and (c) the QHAS curves on
the two azimuths. On $\az{11\bar2}$, panels (a) and (b), the rates trace the single cosine
of $\mathbf L_1$ and $\mathbf L_2$, with a maximum near $\bar{\rm M}$. On $\az{1\bar10}$,
panel (c), the three directions give two cosines and a double maximum. These features are
fixed by the lattice and confirm NN jumps between fcc sites~\cite{Jardine2010,Graham1999}.
Curves at different temperatures, coverages and isotopes collapse onto the same shape. The
ratio of a measured rate to $\bar\alpha[1-f_{\K}]$ lies between about $0.8$ and $1.4$, and
it departs most from one at the smallest momentum transfers. The node rate therefore
carries most of the information of a curve. The small remaining deformation is the tilt.

\begin{table}[!t]
\centering
\caption{Momentum-resolved curves and their fits with \eqref{eq:shape}. The HSE curves
are from Ref.~\cite[Fig.~2]{Jardine2010}, the QHAS curves on $\az{11\bar2}$ and
$\az{1\bar10}$ from Ref.~\cite[Figs.~7 and 8]{Graham1999}. $n$ is the number of points.
$\bar\alpha$ is the node rate and $\delta$ the tilt, both fitted to the measured rates. The column KMC gives the simulated tilt,
with the coupling fitted in Sec.~5.2. $g_0$ is the value of the
interaction factor $g$ at the nodes, simulated at $0.1$ and $0.2$~ML. At $0.05$ and
$0.15$~ML ($^{a}$), where $|\beta\varepsilon|\le0.06$, it is calculated from \eqref{eq:gc}, and the memory changes it by less than $0.01$. $\Gamma_0=\bar\alpha/g_0$ is the
extracted single-adsorbate total hop rate, the measured node rate divided by the simulated, or calculated ($^{a}$), $g_0$. Fitted are $\bar\alpha$ and $\delta$ of each curve.}
\label{tab:kfit}
\vspace{4pt}
\footnotesize
\begin{tabular}{l c c c c c c c c c c}
\hline
Data & Isotope & $T$ (K) & $\theta$ & Azimuth & $n$ & $\bar\alpha$ (ns$^{-1}$) & $\delta$ & KMC & $g_0$ & $\Gamma_0$ (ns$^{-1}$)\\
\hline
HSE & H & $220$ & $0.10$ & $\az{11\bar2}$ & $26$ & $141.9\pm3.2$ & $-0.40\pm0.08$ & $+0.01$ & $1.00$ & $141.7$\\
HSE & H & $140$ & $0.10$ & $\az{11\bar2}$ & $14$ & $29.0\pm0.5$ & $-0.42\pm0.05$ & $-0.12$ & $1.09$ & $26.6$\\
HSE & D & $140$ & $0.10$ & $\az{11\bar2}$ & $13$ & $15.4\pm0.5$ & $-0.36\pm0.09$ & $-0.12$ & $1.09$ & $14.1$\\
HSE & H & $90$ & $0.10$ & $\az{11\bar2}$ & $22$ & $3.3\pm0.1$ & $-0.36\pm0.15$ & $-0.22$ & $1.16$ & $2.85$\\
QHAS & H & $250$ & $0.05$ & $\az{11\bar2}$ & $16$ & $159.5\pm3.9$ & $-0.15\pm0.07$ & $+0.01$ & $0.98^{a}$ & $162.3$\\
QHAS & H & $200$ & $0.10$ & $\az{11\bar2}$ & $35$ & $92.0\pm1.4$ & $-0.22\pm0.05$ & $-0.04$ & $1.02$ & $90.3$\\
QHAS & H & $170$ & $0.10$ & $\az{11\bar2}$ & $26$ & $55.9\pm1.1$ & $-0.30\pm0.07$ & $-0.09$ & $1.05$ & $53.2$\\
QHAS & H & $140$ & $0.10$ & $\az{11\bar2}$ & $30$ & $25.2\pm0.9$ & $-0.36\pm0.14$ & $-0.12$ & $1.09$ & $23.1$\\
QHAS & H & $250$ & $0.10$ & $\az{1\bar10}$ & $27$ & $221.8\pm2.9$ & $-0.17\pm0.04$ & $+0.01$ & $0.98$ & $226.6$\\
QHAS & H & $200$ & $0.15$ & $\az{1\bar10}$ & $29$ & $105.1\pm1.8$ & $-0.06\pm0.06$ & $-0.04$ & $1.05^{a}$ & $100.4$\\
QHAS & H & $160$ & $0.20$ & $\az{1\bar10}$ & $16$ & $69.9\pm1.8$ & $-0.15\pm0.08$ & $-0.17$ & $1.18$ & $59.1$\\
\hline
\end{tabular}
\end{table}

Each curve is fitted with \eqref{eq:shape}. The fit is a weighted linear least-squares
fit in the two unknowns $\bar\alpha$ and $\bar\alpha\delta$, with the digitized error bars
as weights (Appendix~\ref{app:chi2}). The tilt is the ratio of the two, and its error is
obtained from their covariance. No model of the layer enters this fit.
Table~\ref{tab:kfit} collects the results. The node rate $\bar\alpha$ is fixed to a few
per cent, by all the points of a curve. The tilt is fixed with errors of $0.04$ to $0.15$,
for values of $-0.06$ to $-0.42$.

The node rates give the hop rates. The HSE node rates differ from the published total hop
rates of Ref.~\cite{Jardine2010} by at most $8$\%. At $140$~K the QHAS node rate is
$0.87$ times the HSE node rate. The two techniques thus give nearly the same hop rate.
Between $220$ and $140$~K the HSE node rates give an activation energy of $53$~meV, and
between $200$ and $140$~K the QHAS node rates give $52$~meV. The hop rates fitted to the
QHAS coverage series in Sec.~5.2 are lower at $140$~K, by a factor of about $2.3$, since
the coverage series has a steeper Arrhenius law than the node rates. The origin of
this difference is not resolved here. The
coverage series and the momentum-resolved curves were measured in different runs and
with different incident energies, and a difference in the calibration of the widths is
possible. It does not affect the coupling, which depends only on the coverage dependence
of the series, but it sets the scale of the diffusion coefficients
(Appendix~\ref{app:diff}). The extracted hop
rate is $\Gamma_0=\bar\alpha/g_0$, with the simulated $g_0$ of Sec.~5.3. At $0.1$~ML,
$g_0$ lies between $0.98$ and $1.16$ in Table~\ref{tab:kfit}. The extracted hop rates thus
differ from those of a single-adsorbate analysis, with $g_0=1$, by at most this factor.
They contain the friction of the substrate, and $g_0$ contains none.

One difference lies outside the lattice gas. At $250$~K the QHAS node rate on
$\az{1\bar10}$ is $1.4$ times that on $\az{11\bar2}$, whereas the two azimuths must give
the same hop rate. The curves were measured with different incident energies, and the
difference is comparable to the spread reported between data sets~\cite{Graham1999}. At
$200$~K the ratio of the extracted hop rates on the two azimuths is $1.1$.

The tilt is the only property of the layer that a single curve contains. It is, however,
much harder to extract than the node rate, for four reasons. First, it is a small
deformation. Over a whole curve it changes the rate by $10$ to $40$\%, and at $0.15$ and
$0.2$~ML the predicted tilts are only half an error bar to two error bars from
zero. Second, it is fixed
mainly by the rates at long wavelength, where $1-f_{\K}$ is small and the relative errors
are largest. Removing the two lowest-$K$ points of an HSE curve changes its tilt by up to
$0.09$ (Appendix~\ref{app:robust}). Third, a constant added to the Chudley--Elliott form
fits the HSE curves as well as a tilt (Appendix~\ref{app:robust}). A single curve cannot
separate the two. Fourth, all curves are digitized, and the QHAS error bars are likely
underestimated (Appendix~\ref{app:digit}). The tilt is thus a meaningful observable only
for curves that extend to small momentum transfers with reliable error bars.

With these limits, the tilts at $0.15$ and $0.2$~ML are consistent with the coupling of
Sec.~5.2. The measured values, $-0.06\pm0.06$ and $-0.15\pm0.08$, lie $0.3$ and $0.25$ standard deviations from the simulated $-0.04$ and $-0.17$. A constant coupling of
$-8.6$~meV would give $-0.25$ and $-0.40$, about $3.1$ to $3.2$ standard deviations from the measured
values. With the digitized errors the constant coupling is therefore disfavoured.
If the true errors were $1.6$ times larger, the difference would fall to about two standard
deviations. The tilts at these two coverages thus support the coupling of the coverage
series but do not establish it by themselves.

The two tilts can be added to the fit of the coupling, with the simulated tilts of
Appendix~\ref{app:kmctilt} as the model. Fitted together with the coverage series, they
give $e_p=9.4\pm1.5$~meV, $T^*=224\pm20$~K and $s_p=0.49\pm0.10\,k_B$, with $\chi^2=2.4$
for $12$ values and $8$ degrees of freedom. These values agree with those of the coverage
series alone. The narrower $T^*$ rests on the two tilts and on their digitized errors.
Since the tilts do not depend on $\Gamma_0$, their agreement with $\varepsilon(T)$ favours
the pair entropy, rather than a friction of the substrate growing with the coverage, as
the origin of the fall of the prefactor. The quantities calculated in this work use
$e_p=9.5$~meV and $T^*=222$~K, which lie within these errors.

\subsection{The shape of the curves at low coverage and the isotope effect}

This subsection examines the shape of the measured curves at low coverage and the isotope
effect. The HSE data contain curves of H and D at the same temperature, $140$~K, so that
both questions are answered with the same figure.

\begin{figure}[!t]
\centering
\includegraphics[width=0.99\textwidth]{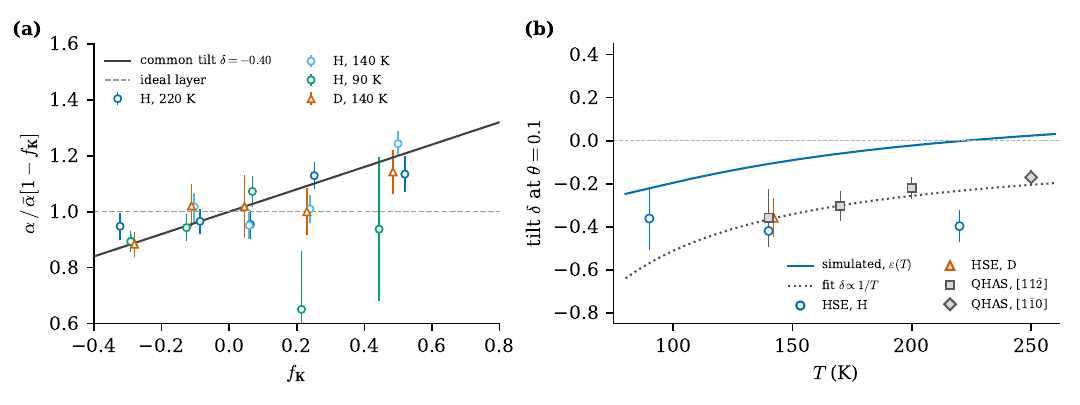}
\caption{Shape of the measured dephasing rate at $0.1$~ML for H and D. (a) Ratio of the
measured HSE rate $\alpha$ to the Chudley--Elliott form $\bar\alpha[1-f_{\K}]$ against
$f_{\K}$ on $\az{11\bar2}$, for H at $220$, $140$ and $90$~K and D at
$140$~K~\cite{Jardine2010}. Each symbol is the weighted mean of the measured ratios in a
bin of $f_{\K}$, and the symbols of different curves are shifted slightly in $f_{\K}$ for
clarity. Points with $1-f_{\K}<0.25$ are omitted. (b) Tilt $\delta$ against temperature at
$0.1$~ML, for the HSE curves of H (circles) and D (triangle) and the QHAS curves of H on
$\az{11\bar2}$ (squares) and $\az{1\bar10}$ (diamond). In (a) the solid line is the common
tilt $\delta=-0.40$, the weighted mean of the tilts of the four measured curves, and the dashed line is the
calculated ideal layer. In (b) the symbols are the tilts fitted to the measured curves
(Table~\ref{tab:kfit}). The solid line is the simulated tilt for the
fitted coupling (Appendix~\ref{app:kmctilt}). The dotted
line is the law $\delta\propto1/T$ fitted to the measured tilts. Error bars in (a) are
those of the bin means, from the published errors. Error bars in (b) are the standard
errors of the fits.}
\label{fig:shape}
\end{figure}

Figure~\ref{fig:shape} shows that the measured curves at $0.1$~ML are tilted more than the
coupling predicts, and that H and D share the same shape. Panel (a) displays the ratio of
the measured HSE rates to the Chudley--Elliott form against $f_{\K}$. On this plot the
ideal layer gives a horizontal line at one, and a tilt $\delta$ gives a straight line of
slope $-\delta$. Panel (b) displays the tilt at $0.1$~ML against temperature, measured and
simulated.

In panel (a) the four HSE curves lie on one straight line, with the common tilt
$\delta=-0.40$. The ratio rises from about $0.9$ near the zone boundary, where $f_{\K}<0$,
to about $1.2$ at long wavelength, where $f_{\K}\approx0.5$. The curve at $90$~K scatters
most, since its rates are the smallest. The curves of H and D at $140$~K lie on top of
each other. Their tilts are $-0.42\pm0.05$ and $-0.36\pm0.09$. The deformation is therefore
common to both isotopes. It also survives the removal of the lowest-$K$ points and of the
points beyond $\bar{\rm M}$ (Appendix~\ref{app:robust}).

In panel (b) the measured HSE tilts stay near $-0.40$ from $90$ to $220$~K. The QHAS
tilts at $0.1$~ML lie between $-0.17$ and $-0.36$. The QHAS curve at $0.05$~ML and
$250$~K, with $-0.15\pm0.07$ against a simulated $+0.01$, shows the same excess. The simulated tilt for the fitted
coupling, the solid line, is $-0.22$ at $90$~K, $-0.12$ at $140$~K and about zero at $220$~K.
The measured tilts thus exceed it in both techniques, and the excess grows with the
temperature. At $0.15$ and $0.2$~ML no excess appears (Sec.~5.4). The symmetric jump rule of Sec.~5.3
raises the simulated tilt at low temperature, to $-0.35$ at $80$~K, and so reduces the
excess there, but not near $220$~K, where the coupling vanishes. The excess is not a sign
of longer jumps, which would move the maximum away from $\bar{\rm M}$.

Several mechanisms were tested against this excess. A repulsion of any range gives
$\delta\propto\theta(1-\theta)/T$. Fitted to the measured tilts at $0.1$~ML, the dotted
line, it gives $\chi^2=8.8$ for $7$ degrees of freedom. It predicts, however, $-0.36$ and
$-0.57$ at $0.15$ and $0.2$~ML, about five standard deviations from the measured values
with the digitized errors. The momentum dependence of the collective friction is
contained in the simulated tilts and fails with them. Jumps through the hcp sites give a
ratio that grows towards the zone boundary (Appendix~\ref{app:hcp}), opposite to panel
(a). The lattice-gas mechanisms are thus excluded by the temperature dependence, the
coverage dependence or the sign of the excess. Both techniques show the excess, so that
an instrumental effect is unlikely. A decay channel independent of $\K$ that weakens with
coverage remains possible, for instance one tied to a fixed number of defects or steps.
It adds a constant to the dephasing rate, of $0.11$ to $0.38$ times $\bar\alpha$ for the
HSE curves (Appendix~\ref{app:robust}). Since a tilt is hard to extract (Sec.~5.4), part
of the excess may also come from the digitized long-wavelength points. Testing these
possibilities requires the full decays at long wavelength, which are not published. The
hop rates are not affected, since they are fixed by the rate at the nodes. The excess is
itself a result. It points to physics outside the NN lattice gas at low coverage, which a
single-adsorbate analysis, with no tilt, could not detect.

The isotopes differ mainly in the single-adsorbate hop rate. The interaction factor $g$
depends only on $\theta$, $\K$ and $\beta\varepsilon$. The pair energy is electronic and
common to both isotopes, and in the classical limit the pair entropy \eqref{eq:sp} depends
only on frequency ratios, which are also common. The zero-point part of the vibrational
free energy \eqref{eq:fvib}, however, scales as the inverse square root of the mass. Take the two parallel modes of
each adparticle at $27.7$~meV for H and $19.6$~meV for D, lowered by $7$ to $12$\% by an
occupied neighbour, the range that gives $s_p$ between $0.27$ and $0.50\,k_B$ for four modes. The quantum $\Delta f_{\rm vib}$ of \eqref{eq:fvib} then makes the coupling of D more repulsive than that of H by $0.6$ to
$1.0$~meV at $140$~K. This shifts $\beta\varepsilon$ by $-0.05$ to $-0.09$ and raises $g_0$
of D by $2$ to $3$\% at $0.1$~ML, less than the errors of the node rates. The interaction
factor is therefore common to both isotopes within the errors. The friction of the
substrate differs between H and D, but it enters only $\Gamma_0$. The collective friction
depends on the coupling and is common to both. The same shape of H and D in panel (a)
agrees with this. At $140$~K the measured node rates give
$\bar\alpha_{\rm H}/\bar\alpha_{\rm D}=1.88\pm0.07$. Since $g_0$ is common, this is also the
ratio of the extracted hop rates. The published Arrhenius rates of Ref.~\cite{Jardine2010}
give a ratio of $1.0$--$1.4$ above $200$~K, about $1.4$ between $150$ and $105$~K, and
$1.8$--$2.0$ below $95$~K. The node rates come from the momentum-resolved curves of the same reference. The two
published data sets thus differ near $140$~K. The rise of the ratio at low temperature is the isotope effect of tunnelling. It
belongs to the single adsorbate and needs no correction for the interaction. Once the
interaction is divided out, the QHAS widths of the coverage series give activation
energies of $85\pm3$~meV for H and $83\pm3$~meV for D, against $52$~meV from the node
rates (Sec.~5.4). The higher value reported for D~\cite{Graham1999} thus does not require a
different barrier, in line with Ref.~\cite{Jardine2010}. The collective diffusion
coefficients of the fitted model are given in Appendix~\ref{app:diff}. Since $\Pi$ and $R$
are common to both isotopes, their ratio for H and D equals that of the single-adsorbate
coefficients at every coverage.

\section{Comparison with standard quantum transition-state theory}

The previous section converted the measured rates into extracted single-adsorbate hop
rates, $\Gamma_0=\bar\alpha/g_0$. This section compares them with standard quantum
transition-state theory (TST), which calculates $\Gamma_0$ for one adsorbate. The comparison
carries three messages, one per subsection after the first, which recalls the theory.
First, the extracted hop rates give the physical parameters of the adsorbate--substrate
system, in particular the friction of the substrate. Second, the memory equation has the
structure of a TST for the adlayer. Third, the two theories combine into one rate of the
adlayer.

\subsection{Standard quantum TST of a single adsorbate}

Standard quantum TST holds above the
crossover~\cite{TorresMiyares2025,Hanggi1990,Wolynes1981}. It gives the rate of the jumps
of one adsorbate along one symmetry direction, that is, along one pair $\pm\mathbf L_j$.
A measurement on one azimuth sees only the equivalent directions that are not orthogonal
to $\K$. Let $n_{\rm dir}$ be their number. On Pt(111) $n_{\rm dir}=2$ on $\az{11\bar2}$ and
$n_{\rm dir}=3$ on $\az{1\bar10}$, and its largest value, $3$, counts all the directions. The
hop rate seen on the azimuth is the sum over these directions,
\begin{equation}
\Gamma_{\rm dir}\;=\;n_{\rm dir}\,\frac{\omega_0}{2\pi}\,\frac{\lambda^\ddagger}{\omega^\ddagger}\,
\Xi\;e^{-\beta V^\ddagger} .
\label{eq:tst}
\end{equation}
Here $V^\ddagger$ is the barrier height, and $\omega_0$ and $\omega^\ddagger$ are the well
and barrier frequencies. On $\az{11\bar2}$, $\Gamma_{\rm dir}=\tfrac23\Gamma_0$, the rate
reported by the CF analysis of a single adsorbate~\cite{TorresMiyares2025,TorresMiyares2026a}.
With $n_{\rm dir}=3$, $\Gamma_{\rm dir}=\Gamma_0$, the total hop rate. The Chudley--Elliott
form \eqref{eq:ce} already contains the weighting of the azimuth through $1-f_{\K}$, so
that the extracted $\Gamma_0$ is the total hop rate. It is compared with \eqref{eq:tst} for
$n_{\rm dir}=3$, as in Refs.~\cite{Jardine2010,TorresMiyares2025}, so that the friction
obtained below can be compared directly with those references.

The TST rate is the classical attempt rate $\omega_0/2\pi$ times the Boltzmann factor of
the barrier, corrected in two ways. The dynamical correction is the friction factor
$\lambda^\ddagger/\omega^\ddagger$, with
$\lambda^\ddagger=(\omega^{\ddagger2}+\gamma^2/4)^{1/2}-\gamma/2$ for an Ohmic friction of
coefficient $\gamma$. The friction makes the adsorbate recross the barrier and lowers the
rate. The static correction is the Wolynes factor $\Xi$, the quantum correction of the
barrier and of the well. It is the ratio of the quantum partition functions at the top of
the barrier and at the bottom of the well, each divided by its classical limit
$Q_{\rm cl}(\omega)=k_BT/\hbar\omega$. At the top the unstable mode has the imaginary
frequency $i\omega^\ddagger$. Without friction, with $Q$ of \eqref{eq:Q0},
\begin{equation}
\Xi_0\;=\;\frac{Q(i\omega^\ddagger)/Q_{\rm cl}(i\omega^\ddagger)}{Q(\omega_0)/Q_{\rm cl}(\omega_0)}
\;=\;\frac{\sinh(\beta\hbar\omega_0/2)}{\beta\hbar\omega_0/2}\;
\frac{\beta\hbar\omega^\ddagger/2}{\sin(\beta\hbar\omega^\ddagger/2)} .
\label{eq:Xi0}
\end{equation}
The first factor is the correction of the well, which contains its zero-point energy.
The second is the correction of the barrier, which contains the tunnelling through its
top. With Ohmic friction both become products over the Matsubara frequencies
(Appendix~\ref{app:tst}), and $\Xi$ diverges at the crossover temperature
$\hbar\lambda^\ddagger/2\pi k_B$. The TST does not remove this divergence. It holds only
well above the crossover and cannot describe the passage to deep tunnelling.

The barrier is specified as in the stochastic-wave-function solutions of
Ref.~\cite{TorresMiyares2025}. The corrugation along a jump is a cosine,
$V(x)=\tfrac12V^\ddagger[1+\cos(2\pi x/a)]$, with $x$ the position along the jump counted
from the top of the barrier. Its well and barrier curvatures are equal, and both
frequencies are fixed by the height,
\begin{equation}
\hbar\omega_0\;=\;\hbar\omega^\ddagger\;=\;\frac{2\pi\hbar}{a}\Bigl(\frac{V^\ddagger}{2m}\Bigr)^{1/2} ,
\label{eq:parab}
\end{equation}
with $m$ the mass of the adsorbate. The parameters are those of
Ref.~\cite{TorresMiyares2025}, $V^\ddagger=72$~meV for both isotopes and $\gamma=140$ and
$70$~ps$^{-1}$ for H and D. Equation~\eqref{eq:parab} then gives $\hbar\omega_0=27.7$ and
$19.6$~meV. The anharmonicity of the cosine adds two small corrections. The exact
transmission through the barrier is $2$--$8$\% smaller than that of the harmonic barrier
at $80$--$120$~K, and the exact eigenenergies of the well raise the partition function of
the well by about $10$\%. With $\gamma/\omega^\ddagger\approx3$ the motion is strongly
damped, and the crossover temperature of \eqref{eq:tst} falls to about $14$~K, far below
the $65$~K estimated from the data. The TST is therefore reliable only in the activated
range. The missing tunnelling at low temperature lies in the treatment of the
dissipation, not in the shape of the barrier (Appendix~\ref{app:tst}).

\subsection{Physical parameters of the adsorbate--substrate system}

The first message is a method. Single-adsorbate theories give the barrier, the quantum
corrections and the friction of the substrate only through the hop rate of an isolated
adsorbate. The measured rates must therefore first be converted into that hop rate. The
conversion is made at the nodes, $\Gamma_0=\bar\alpha/g_0$, with the simulated $g_0$ of
Sec.~5.3. There the SSF nearly equals $1-\theta$ whatever the coupling, so that $g_0$
depends little on the structure of the layer. The HSE Arrhenius plot was moreover
measured practically at a node.

\begin{figure}[!t]
\centering
\includegraphics[width=0.99\textwidth]{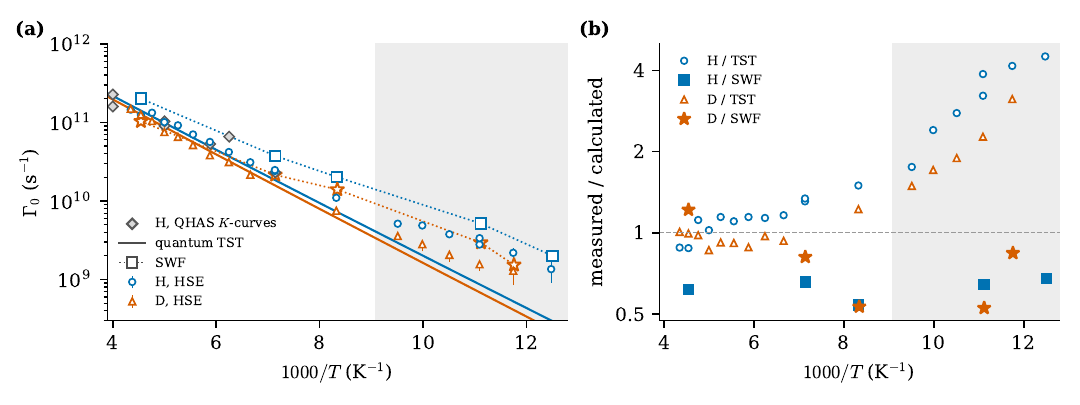}
\caption{Extracted single-adsorbate hop rates compared with two single-adsorbate theories,
for H (blue) and D (orange). (a) Total hop rate against $1000/T$. (b) Ratio of the
extracted hop rate to the TST rate (open symbols) and to the SWF rate (filled symbols)
against $1000/T$. In (a) the small open symbols are the published HSE total hop rates at
$0.1$~ML~\cite{Jardine2010} divided by the simulated $g_0$ of Fig.~\ref{fig:memory}(a).
The grey diamonds are $\bar\alpha/g_0$ from the QHAS curves of Table~\ref{tab:kfit}. Both are
extracted. The lines are the standard quantum TST \eqref{eq:tst} for $n_{\rm dir}=3$, calculated with the cosine
corrugation, $V^\ddagger=72$~meV and $\gamma=140$ and $70$~ps$^{-1}$, with nothing fitted.
The large open symbols with dotted lines are the stochastic-wave-function (SWF) rates of
the isolated adsorbate, calculated in Ref.~\cite{TorresMiyares2025} and taken from its
Fig.~1. They are the rates of the $n_{\rm dir}=2$ directions seen on $\az{11\bar2}$ and are
multiplied by $3/n_{\rm dir}=3/2$ to give total hop rates. In (b) the shaded range below
$110$~K is where the TST fails.
Error bars in (a) are those of Ref.~\cite{Jardine2010}, divided by $g_0$.}
\label{fig:tst}
\end{figure}

Figure~\ref{fig:tst} compares the extracted hop rates with two single-adsorbate theories,
the standard quantum TST and the stochastic wave functions. Panel (a) displays the total
hop rate against $1000/T$, extracted and calculated. Panel (b) displays the ratio of the
extracted rate to each theory. No parameter is fitted in the figure.

Panel (a) shows that the two techniques give the same hop rate. Where the HSE and QHAS
rates overlap, they differ by less than $30$\%. The HSE rates bend away from a straight
Arrhenius line below about $110$~K, the mark of tunnelling.

Panel (b) shows where each theory works. Between $140$ and $230$~K the ratio of the
extracted rate to the TST rate lies between $0.86$ and $1.01$ for D and between $0.88$
and $1.34$ for H. Below $110$~K the TST falls short, and the ratio reaches $4.5$ for H at
$80$~K and $3.1$ for D at $85$~K. The reason is the crossover temperature of the TST,
about $14$~K, far below the $65$~K of the data. The SWF rates reproduce the bending of the
Arrhenius plot down to $80$~K. For H the ratio stays between $0.54$ and $0.68$, a nearly
constant factor whose origin is not resolved here. At and below $140$~K the simulated $g_0$ lies
between $1.09$ and $1.17$ at $0.1$~ML, and between $0.97$ and $1.03$ with the symmetric jump
rule of Sec.~5.3. Dividing it out changes the extracted rates by at
most this factor and does not change these conclusions.

The friction of the substrate is obtained from the same comparison. In the strongly damped
regime $\lambda^\ddagger/\omega^\ddagger\simeq\omega^\ddagger/\gamma$, so that the inferred
friction scales as the inverse of the hop rate. A fit of \eqref{eq:tst} to the extracted
HSE hop rates above $140$~K, with $V^\ddagger$ and the frequencies fixed, gives
$\gamma=124$~ps$^{-1}$ ($119$--$128$) for H and $75$~ps$^{-1}$ ($72$--$77$) for D. These
values are $0.89$ and $1.07$ times those of Ref.~\cite{TorresMiyares2025}. The fit is good
for D, with $\chi^2=3.1$ for $8$ degrees of freedom, and poorer for H, with $\chi^2=46.9$
for $10$ (Appendix~\ref{app:chi2}). The poor fit of H does not come only from the lowest temperatures. Restricted to
$170$--$230$~K, the fit gives $131$~ps$^{-1}$ with $\chi^2=23.0$ for $6$ degrees of freedom.
The H rates scatter more than their published errors. The rate at $220$~K, for instance,
lies below that at $210$~K. The fitted friction also rises slowly as the range is
restricted, to $143$~ps$^{-1}$ above $200$~K. This points to a temperature dependence of the
TST slightly different from the measured one for H, whereas D is closer to the classical
limit.
A single-adsorbate analysis, with $g_0=1$, would underestimate the friction by the factor
$g_0$. This bias is small at $0.1$~ML, where $g_0=0.98$--$1.09$ at and above $140$~K. It
grows with the coverage, to $1.31$ at $0.2$~ML and $1.91$ at $0.4$~ML at $135$~K
(Table~\ref{tab:rule}).

The friction of the substrate does not depend on the coverage. This is tested with the
coverage series of Sec.~5.2, by letting the prefactor of the single-adsorbate rate vary
with the coverage in the fit of the measured widths of Fig.~\ref{fig:coverage}(a). Since
$\Gamma_0\propto1/\gamma$ in the strongly damped regime, the fitted prefactors give
$\gamma(0.2)/\gamma(0.1)=0.95\pm0.09$ and $\gamma(0.4)/\gamma(0.1)=1.01\pm0.13$. The
friction of the substrate is thus the same at the three coverages within these errors.

The cosine barrier also gives the pair entropy of Sec.~3. A barrier lowered by $e_p$ is also
softer, and each occupied neighbour lowers both frequencies by about $e_p/(2V^\ddagger)$. With
\eqref{eq:sp} this gives $s_p=0.18$--$0.37\,k_B$, depending on the modes included, against
$0.50\pm0.14\,k_B$ fitted. The softening of the cosine barrier thus accounts for most of the
pair entropy.

\subsection{The memory equation as a TST of the adlayer}

The second message is an analogy between the two theories. Table~\ref{tab:analogy}
summarizes it. The TST rate of one adsorbate is the limit $t\to0^+$ of the flux--side
correlation function, where every crossing of the barrier counts. Later recrossings,
caused by the friction of the substrate, reduce it~\cite{Hanggi1990}. For the adlayer the
same construction applies to the density wave at $\K$. Its initial decay rate $\Omega$
counts every jump. The back jumps caused by the collective friction reduce the rate to
$\lambda_{\min}\le\Omega$. The local rate $\Omega$ is thus the transition-state rate of the
density wave, and $\lambda_{\min}/\Omega$ its transmission coefficient.

\begin{table}[!t]
\centering
\caption{Correspondence between the standard quantum TST of a single adsorbate and the
memory equation of the adlayer.}
\label{tab:analogy}
\vspace{4pt}
\small
\begin{tabular}{p{4.0cm} p{5.0cm} p{5.0cm}}
\hline
 & Single adsorbate & Adlayer\\
\hline
Variable & position $x$ along the jump & density wave $\delta n_{\K}$\\
Transition-state rate & $\omega_0e^{-\beta V^\ddagger}/2\pi$ & local rate $\Omega$\\
Static correction & Wolynes factor $\Xi$ & static factor $\Pi$\\
Dynamical correction & $\lambda^\ddagger/\omega^\ddagger$ & $\lambda_{\min}/\Omega$\\
Friction & friction of the substrate & collective friction $M(\K,\tau)$\\
Friction without memory & Kramers rate, $\gamma$ & $\Omega/R$, $\hat M(\K,0)$\\
Friction with memory & Grote--Hynes frequency $\lambda^\ddagger$ & lowest rate $\lambda_{\min}$, \eqref{eq:pole}\\
Divergence at the crossover & in $\Xi$ & none\\
\hline
\end{tabular}
\end{table}

The analogy holds factor by factor. The static factor $\Pi$ is the counterpart of the
Wolynes factor $\Xi$, in role but not in construction. $\Xi$ compares quantum partition
functions of one adsorbate at the barrier and in the well. $\Pi$ is a classical average
over the configurations of the layer. The transmission coefficient $\lambda_{\min}/\Omega$
is the counterpart of the friction factor $\lambda^\ddagger/\omega^\ddagger$. The regimes
of rate theory also correspond one by one. Without friction the rate is the
transition-state rate, and without collective friction $\lambda_{\min}=\Omega$. For Ohmic
friction the Kramers rate is fixed by the friction at zero frequency~\cite{Kramers1940}. By
\eqref{eq:R}, $1/R$ is likewise fixed by the integral of the collective friction, its value
at zero frequency. $\Omega/R$ is thus the rate for a collective friction without memory.
For a friction with memory, $\hat\gamma(s)$, the Grote--Hynes frequency solves
$\lambda^\ddagger=\omega^{\ddagger2}/[\lambda^\ddagger+\hat\gamma(\lambda^\ddagger)]$, with
the friction evaluated at the reactive frequency~\cite{GroteHynes1980}. The lowest rate of
the adlayer obeys \eqref{eq:pole}, with the collective friction evaluated at the rate
itself. For an exponential friction kernel, $\hat\gamma(s)=\gamma/(1+s\tau_\gamma)$, the
integral $M(\K,0)/\Omega_1$ corresponds to $\gamma$, the memory time $1/\Omega_1$ to
$\tau_\gamma$, and \eqref{eq:lmin} to the Grote--Hynes frequency. The simulated memory time
is $0.3$--$0.5$ times the relaxation time (Sec.~5.1). The adlayer therefore lies in the
Grote--Hynes regime, and its dephasing rate lies below the Kramers-like value $\Omega/R$.

The TST of the adlayer has no divergence. The static factor $\Pi$ is a classical average
over the configurations, with no unstable mode and no imaginary frequency. The lowest rate
is given by the exact memory equation at every temperature. All quantum effects, the
divergence of the Wolynes factor included, stay inside $\Gamma_0$. The analogy extends to the action of the barrier
(Appendix~\ref{app:action}).

The single-adsorbate TST lacks the factors of the adlayer. It sets $g=1$ and fixes the
curve of the dephasing rate to the Chudley--Elliott curve, so that the tilt is invisible
to it. With one barrier for all jumps it cannot give the fall of the prefactor with
coverage, from $6.8$ to $1.5$~THz between $0.1$ and $0.6$~ML~\cite{Graham1999}, nor the
compensation rule. No TST gives the memory, which lies outside the transition-state
assumption. The analogy also has a limit. Coherent tunnelling, in which the adsorbate keeps
its phase and moves as a wave spread over many sites, has no counterpart, since the memory
equation describes jumps. Incoherent tunnelling, in which the adsorbate loses its phase
between transitions, proceeds by jumps and is included. For H on Pt(111) the damping
$m a^2\gamma/2\pi\hbar\approx27$ excludes coherent tunnelling.

\subsection{The rate of the adlayer}

The third message is a combination of the two theories into one rate. Let the hop rate of
the layer be the dephasing rate divided by the geometry of the lattice,
$\Gamma_L(\K)=\alpha(\K)/[1-f_{\K}]$. It is the rate of the layer as the measurement sees
it, not the hop rate of one adsorbate. With $\alpha=\Gamma_0\,g(\K)[1-f_{\K}]$,
$g=\Pi\lambda_{\min}/\Omega$ from \eqref{eq:g} and $\Gamma_0$ from \eqref{eq:tst} with
$n_{\rm dir}=3$,
\begin{equation}
\Gamma_L(\K)\;=\;n_{\rm dir}\,\frac{\omega_0}{2\pi}\,e^{-\beta V^\ddagger}\;
\frac{\lambda^\ddagger}{\omega^\ddagger}\;\Xi\;\Pi(\K)\,\frac{\lambda_{\min}}{\Omega} .
\label{eq:layertst}
\end{equation}
Equation~\eqref{eq:layertst} has the same structure for the adsorbate and for the layer, a
classical rate with a static and a dynamical correction for each. Without
$\lambda_{\min}/\Omega$ the product is the transition-state rate of the density wave, and
with it the dephasing rate. The two frictions act on different variables and on different
time scales, one adsorbate within $1/\gamma\approx7$~fs and the density wave over the time
between jumps. Only their transmission coefficients combine, as the product
$(\lambda^\ddagger/\omega^\ddagger)(\lambda_{\min}/\Omega)$.

The factors are calculated for H at the nodes, with the cosine parameters, the fitted
coupling and the simulated $\lambda_{\min}/\Omega$. They include the lowering of the barrier
by the occupied neighbours through the factor $B(\nu)$ of Appendix~\ref{app:layer}. The static
factor $\Pi_0$ quoted here therefore differs from that of Table~\ref{tab:rule}. At $0.1$~ML the factors of the layer
lie between $0.75$ and $1.25$. At $140$~K, for instance, $\Pi_0=1.11$ and
$\lambda_{\min}/\Omega=0.93$, against $\Xi=1.28$ and $\lambda^\ddagger/\omega^\ddagger=0.28$.
At $0.4$~ML and $100$~K, $\Pi_0=2.72$ and $\lambda_{\min}/\Omega=0.68$. The product of the
two transmission coefficients is then $0.28\times0.68=0.19$. The static factor of the layer
exceeds the Wolynes factor, $1.53$, and is the largest factor that raises the rate. The rate of
the layer, $3.7\times10^9$~s$^{-1}$, is $1.8$ times the TST rate of the isolated adsorbate.
At $230$~K the coupling has nearly vanished, and the factors of the layer lower the rate
slightly, by $5$ to $20$\%. $\Gamma_L$ is a property of the signal. At the nodes it is the
hop rate of an isolated adsorbate corrected for blocking, correlation, neighbourhood and
memory.

\section{Conclusions}

The coherent ISF measured by helium scattering is a CF, the CF of the separation between
adparticles at two times. Its memory equation has been applied to the published
dephasing rates of H and D on Pt(111), measured by HSE and QHAS, with a NN Ising lattice
gas. Incoherent tunnelling obeys detailed balance, so that the equation holds. It
separates the single-adsorbate hop rate, which contains the barrier, the quantum
corrections and the friction of the substrate, from the effects of the adlayer, without a
model of the barrier.

The memory function of the equation is interpreted as a friction, the collective, or
configurational, friction. The adlayer exerts it on its own density fluctuations. Its integral over the
delay fixes the memory measure $R$ of Refs.~\cite{PaperI,PaperII}. Its strength and its
memory time together fix the lowest relaxation rate, which is the dephasing rate. The
simulations give a memory time of $0.3$ to $0.5$ times the relaxation time of the density
wave. At the nodes and below about $100$~K the simulated dynamical factor
$\lambda_{\min}/\Omega$ is $0.77$ to $0.86$ at $0.1$~ML and falls to $0.58$ at
$0.4$~ML. At long wavelength $1/R$ stays between $0.98$ and $1.01$, so that the collective
friction nearly vanishes there. The enhancement of collective diffusion by the repulsion
is therefore static. The collisional friction of the interacting single-adsorbate
approximation was an initial and simple attempt to describe this effect. The friction of
the substrate acts instead on each adsorbate and stays inside the hop rate. The two
frictions act on different variables, so that a total friction has no meaning.

The measured rate belongs to the adlayer. The adlayer does not alter the measurement, but
it alters its meaning. The single-adsorbate hop rate $\Gamma_0$ is extracted from the
measured rate by dividing out the interaction factor $g$. A single-adsorbate analysis sets
$g=1$. How $g$ is obtained depends on what has been measured, and two cases occur. In the
first, the dephasing rate is measured at a single momentum transfer, against temperature
or coverage. This is the case of the HSE Arrhenius plot, measured at a node, and of the
QHAS coverage series. No tilt is then available, and the coupling must come from the
coverage dependence of the rates. Here it is fixed by the activation energies and
prefactors published for the coverage series, together with the two Arrhenius parameters
of $\Gamma_0$. The coupling is the free energy of a NN pair, with a pair energy of
$9.5$~meV and a pair entropy of about half the Boltzmann constant. The pair energy makes a
NN pair costly, and the pair entropy, from the softening of the vibrations, makes it
favourable as the temperature rises. The two balance at $T^*=222$~K, where the coupling
vanishes and the layer is ideal at every coverage. $T^*$ is also the compensation
temperature of the coverage series. This coupling gives $\chi^2=2.3$ for $6$ degrees of
freedom, whereas the usual constant coupling gives $14.7$ for $7$. The constant coupling is thus
disfavoured, although this rests on the published errors. The case for the pair entropy
combines these values, the measured widths, the tilts and an estimate from the barrier. With
the coupling fixed, the lattice gas gives the interaction factor $g$ at the measured
momentum transfer without further fitting. Its static part is calculated with the closed
forms, and its dynamical part $\lambda_{\min}/\Omega$ is simulated. The extracted hop rate
is then $\Gamma_0=\alpha/\{g[1-f_{\K}]\}$. At a node it is $\Gamma_0=\alpha/g_0$, nearly free of
the SSF. Without a coverage series the coupling cannot be fixed, and only the dilute
limit, $g=1$, remains.

In the second case, the dephasing rate is measured against the momentum transfer. Each
curve is then summarized, without a model of the layer, by two numbers from a weighted
linear fit, its node rate $\bar\alpha$ and its tilt $\delta$. The node rate is fixed to a
few per cent and gives the hop rate, $\Gamma_0=\bar\alpha/g_0$, as in the first case. The
tilt depends, within the lattice gas, only on the coupling at the temperature and coverage
of the curve. In principle it can fix the coupling where no coverage series exists, or
test a coupling obtained otherwise. In practice the tilt is hard to extract. It is a small
deformation, fixed mainly by the long-wavelength rates, and degenerate on a single curve
with a constant offset. It needs curves that extend to small momentum transfers with
reliable error bars. Here the measured tilts at $0.15$ and $0.2$~ML are consistent with
the coupling of the coverage series and disfavour the constant coupling, with the
digitized errors. They support the pair entropy against a friction of the substrate that
would grow with the coverage, but do not establish it by themselves.

In both cases the simulated $g_0$ lies between $0.98$ and $1.17$ at $0.1$~ML. The
extracted hop rates thus differ from those of a single-adsorbate analysis by at most this
factor at the lowest coverage studied. Below $130$~K the result depends on how the
occupied neighbours modify a tunnelling jump, which no theory fixes. In the QHAS range
the coverage series requires that they lower the barrier of the departure site. A
symmetric rule, which splits the energy difference between the two sites, cannot
reproduce it. Below $130$~K the two rules bracket $g_0$. At $80$~K it lies between $0.97$
and $1.17$ at $0.1$~ML, between $1.02$ and $1.62$ at $0.2$~ML and between $1.18$ and $4.67$ at
$0.4$~ML. At $0.1$~ML the systematic uncertainty is thus a factor of up to $1.2$. At $140$~K the QHAS node rate is
$0.87$ times the HSE node rate, so that the two techniques give nearly the same hop rate.

The adlayer thus provides the coupling, its static effect on the rate and the collective
friction. What remains after dividing out $g$, the hop rate $\Gamma_0$, belongs to the
single adsorbate. It contains the barrier, the quantum corrections and the friction of
the substrate, and only these are compared with single-adsorbate theories. The standard
quantum TST of Sec.~6 is calculated with a cosine barrier and the parameters of
Ref.~\cite{TorresMiyares2025}, a barrier of $72$~meV and frictions of $140$ and
$70$~ps$^{-1}$ for H and D. Between $140$ and $230$~K, without any fit, the ratio of the
extracted to the calculated hop rate lies between $0.86$ and $1.01$ for D and between
$0.88$ and $1.34$ for H. The TST fails below $110$~K, whereas the stochastic-wave-function
rates of Ref.~\cite{TorresMiyares2025} reproduce the bending of the Arrhenius plot down to
$80$~K. With the barrier and the frequencies fixed, a fit of the friction alone to the
extracted hop rates above $140$~K gives $124$ and $75$~ps$^{-1}$ for H and D. The same
division applied to the coverage series gives $\gamma(0.2)/\gamma(0.1)=0.95\pm0.09$ and
$\gamma(0.4)/\gamma(0.1)=1.01\pm0.13$. The friction of the substrate thus does not depend
on the coverage.

At $0.05$ and $0.1$~ML the measured tilts, near $-0.4$ in HSE, exceed the simulated ones
in both techniques. The excess is common to H and D. No lattice-gas mechanism produces it.
A decay channel independent of the momentum transfer, or the digitized long-wavelength
points, may explain it, and its origin is left open. The hop rates at the nodes are not
affected.

The memory equation provides a TST for the adlayer, with the same structure as the
standard quantum TST of a single adsorbate. The static factor of the layer is the
counterpart of the Wolynes factor, and the collective friction plays the role of the
friction of the substrate. The layer lies in the Grote--Hynes regime, and its dephasing
rate lies below the Kramers-like value $\Omega/R$. Unlike the TST of one adsorbate, the TST
of the adlayer has no divergence at the crossover. It is complementary to the TST of one
adsorbate and leaves to it only the hop rate. The analysis is well founded above about
$130$~K, in the QHAS range, and gives an estimate between $130$ and $80$~K
(Appendix~\ref{app:scope}).

Three measurements would test the analysis. HSE at $0.2$--$0.4$~ML below $130$~K, on
both azimuths, would test the temperature-dependent coupling and decide the jump rule for
tunnelling. At the node, at $0.4$~ML and $80$~K, the two rules give $g_0=4.67$ and $1.18$. The same measurement on both isotopes would
test the isotope ratio at finite coverage. HSE curves at $0.05$--$0.4$~ML down to the
smallest momentum transfers, with reliable error bars and the full decays analysed, would
measure the tilt properly and show whether the excess tilt comes from an extra decay
channel. The full decays would also give the collective friction directly, through its
explicit expression.

The approach is well suited to tunnelling diffusion. The memory equation needs only
incoherent jumps that obey detailed balance, and incoherent tunnelling is of this kind.
It needs no model of the barrier and no theory of tunnelling. Whatever theory gives the
hop rate of one adsorbate, standard quantum TST above the crossover or stochastic wave
functions and Grabert--Weiss rates below it, the adlayer only multiplies it by the
interaction factor $g$. Measurements at finite coverage can thus be compared directly with
any single-adsorbate theory of tunnelling. Two limits remain. Coherent tunnelling, in which
the adsorbate keeps its phase over many sites, lies outside the approach, and for H on
Pt(111) the friction of the substrate excludes it. Deep in the tunnelling regime, the
dependence of the jump rates on the occupied neighbours is not fixed by any theory, and
the interaction factor $g$ is then an estimate. A theory of incoherent tunnelling in a
crowded environment would remove this limit.

\appendix
\section{Data and additional analyses}\label{app:B}

This appendix collects the simulated memory function and tilts, the diffusion coefficients
of the fitted model, the least-squares fits, the digitization of the data with the checks
of the excess tilt, the calculation for jumps through the hcp
sites, the parameters and limits of the standard quantum TST, the rate of the adlayer with
neighbour-dependent barriers, the spectral action and the scope of the analysis.

\subsection{Simulated memory function and tilts}\label{app:kernel}

The simulations are described in Sec.~5.1. Table~\ref{tab:kernel} gives the results.

\begin{table}[!h]
\centering
\caption{Memory function at the node $K_0$ of $\az{11\bar2}$. All values are simulated by KMC,
except the column \eqref{eq:lmin}, calculated from the simulated $M(\K,0)$ and $\Omega_1$. Rates are in units
of $\Gamma_0$, the rate of the ideal layer at the node. $\tau_c^{-1}=\Omega/R$ is the
inverse correlation time. The column \eqref{eq:lmin} gives the second-order estimate of
$\lambda_{\min}$. $1/\Omega_1$ is the memory time of the memory function, and
$\hat M(\K,0)/\Omega=1-1/R$ the fraction of $\Omega$ removed by its integral. The
statistical error of $\lambda_{\min}$ is about $2$\%. In the first row, for the ideal layer,
$\lambda_{\min}$ exceeds $\tau_c^{-1}$ within this error.}
\label{tab:kernel}
\vspace{4pt}
\small
\begin{tabular}{c c c c c c c c}
\hline
$\theta$ & $\beta\varepsilon$ & $\Omega$ & $\tau_c^{-1}$ & $\lambda_{\min}$ & Eq.~\eqref{eq:lmin} & $\Omega_1/\lambda_{\min}$ & $\hat M(\K,0)/\Omega$\\
\hline
$0.1$ & $0$ & $1.000$ & $1.002$ & $1.009$ & -- & -- & $0.00$\\
$0.1$ & $-0.60$ & $1.349$ & $1.235$ & $1.127$ & $1.133$ & $2.1$ & $0.085$\\
$0.1$ & $-0.88$ & $1.503$ & $1.272$ & $1.128$ & $1.160$ & $3.1$ & $0.154$\\
$0.4$ & $-0.60$ & $4.108$ & $3.455$ & $2.775$ & $2.904$ & $2.2$ & $0.159$\\
$0.4$ & $-0.88$ & $8.146$ & $5.982$ & $4.433$ & $5.041$ & $3.4$ & $0.266$\\
\hline
\end{tabular}
\end{table}

Without interaction the memory function vanishes within the statistical error. The estimate
\eqref{eq:lmin} deviates most for the strongest coupling, where the memory function has more than
one time scale. The values of $R=\Omega\tau_c$ are $1.09$ and $1.18$ at $0.1$~ML and
$1.19$ and $1.36$ at $0.4$~ML, as the last column gives by \eqref{eq:R}.
A single exponential fitted over the window of the experiments, between $0.3$ and $0.05$,
is $1.03$--$1.05$ times $\lambda_{\min}$. The node values of $g_0$ and
$\lambda_{\min}/\Omega$ quoted in Sec.~5.3 come from such window fits in separate runs and
therefore exceed those of the table by this amount.

\label{app:kmctilt}

The simulated tilts are obtained as described in Sec.~5.1. At $\theta=0.1$ on
$\az{11\bar2}$ the KMC gives $\delta=-0.08\pm0.02$, $-0.13$, $-0.18$, $-0.27$ and $-0.30$ at $\beta\varepsilon=-0.09$,
$-0.32$, $-0.57$, $-0.88$ and $-1.40$. The RPA gives $-0.05$, $-0.17$, $-0.31$, $-0.47$
and $-0.76$. At $\beta\varepsilon=0$ the KMC gives $\delta=0.02\pm0.02$, as expected for the
ideal layer. The KMC values at
$\theta=0.1$ are represented by $\delta=0.416\,\beta\varepsilon+0.153\,(\beta\varepsilon)^2$,
with a largest deviation of $0.04$, at $\beta\varepsilon=-0.09$. At strong coupling the tilt of the window fits is
smaller than the tilt obtained from $\tau_c^{-1}=\Omega/R$, $-0.30$ against $-0.39$ at $\beta\varepsilon=-1.4$. The
difference is systematic, well beyond the statistical errors. At $\theta=0.2$ on $\az{1\bar10}$ the KMC gives
$\delta=-0.19$ and $-0.37$ at $\beta\varepsilon=-0.21$ and $-0.57$. In Table~\ref{tab:kfit} and in the joint fit the simulated tilts at $0.1$ and $0.2$~ML are
interpolated linearly between the simulated couplings. Between $\theta=0.1$ and $0.2$ the
ratio of the simulated tilt to the RPA tilt is interpolated linearly in $\theta$. The
quadratic representation is used only for the line of Fig.~\ref{fig:shape}(b).

\subsection{Diffusion coefficients}\label{app:diff}

The dynamical factor at long wavelength, $1/R(\mathbf 0)$ of \eqref{eq:Dc}, is obtained as described in
Sec.~5.1. At $140$~K,
$\beta\varepsilon=-0.29$, it is $1.012$, $0.996$, $1.013$ and $1.002$ at $0.1$, $0.2$,
$0.4$ and $0.6$~ML. At $230$~K, $\beta\varepsilon=+0.02$, it is $1.008$, $0.997$,
$0.996$ and $0.989$. The statistical error is about $0.008$. For the ideal layer it is
one within the same error. At $\beta\varepsilon=-0.88$ it is $0.98\pm0.01$ at
$0.2$~ML and $1.00\pm0.01$ at $0.4$~ML, the latter from four shorter runs. These are the values quoted in Sec.~5.1. The collective friction therefore changes the collective coefficients of
Table~\ref{tab:D} by less than $2$\%.

The fitted model gives the diffusion coefficients against coverage at $140$ and
$230$~K (Table~\ref{tab:D}). $\Gamma_0$ comes from the single-adsorbate hop rates
fitted to the QHAS widths~\cite{Graham1999}. For H these are the widths of the
coverage series, and for D the widths at $0.1$~ML of the same reference. $F(\mathbf 0)$ comes from the
RPA, and $R(\mathbf 0)$ is set to one. At $230$~K the layer is nearly ideal, and $D_c/D_K$
lies between $0.89$ and $1.00$ up to $0.6$~ML. At $140$~K the repulsion makes the layer less
compressible. $D_c$ then rises to $2.7\,D_K$ at $0.4$~ML and $3.7\,D_K$ at $0.6$~ML, for
both isotopes.

At and above $140$~K the hop rates fitted to the QHAS coverage series and the published HSE hop
rates differ by up to a factor of $2.3$. The QHAS hop rates are higher at
$230$~K and steeper in temperature. This factor is the same at every coverage.
It leaves the coverage analysis unchanged but sets the scale of the table. With the HSE hop rates~\cite{Jardine2010},
$D_K$ would be $0.8$ times the tabulated value at $230$~K and $2.3$ times it at $140$~K,
$5.6\times10^{-5}$ and $1.0\times10^{-5}$~cm$^2$\,s$^{-1}$ for H. These are the values of
Ref.~\cite{TorresMiyares2026a}. With the QHAS hop rates the tabulated value for H at $140$~K,
$4.4\times10^{-6}$~cm$^2$\,s$^{-1}$, is close to the value of about
$5\times10^{-6}$~cm$^2$\,s$^{-1}$ reported in Ref.~\cite{Graham1999}.
The QHAS hop rates also give an isotope ratio of only $1.2$ at $140$~K. The isotope
content of the table is therefore that of the QHAS coverage series.

\begin{table}[!t]
\centering
\caption{Single-adsorbate coefficient $D_K$ on $\az{11\bar2}$, \eqref{eq:DK}, and collective coefficient $D_c$, in
cm$^2$\,s$^{-1}$, for H and D, from the lattice gas with the temperature-dependent
coupling and the single-adsorbate hop rates fitted to the QHAS widths of
Ref.~\cite{Graham1999}. The coefficients are defined by $\avg{L_\parallel^2}=D_Kt$ and
\eqref{eq:einstein}. They are calculated with \eqref{eq:DK} and \eqref{eq:Dc}, with $F(\mathbf 0)$
from the RPA and $R(\mathbf 0)=1$, as the simulations give.}
\label{tab:D}
\vspace{4pt}
\small
\begin{tabular}{c c c c c c c}
\hline
$T$ (K) & & $D_K$ & \multicolumn{4}{c}{$D_c$}\\
 & & & $\theta=0.1$ & $0.2$ & $0.4$ & $0.6$\\
\hline
$140$ & H & $4.4\times10^{-6}$ & $6.0\times10^{-6}$ & $7.8\times10^{-6}$ & $1.2\times10^{-5}$ & $1.6\times10^{-5}$\\
      & D & $3.6\times10^{-6}$ & $5.0\times10^{-6}$ & $6.4\times10^{-6}$ & $9.8\times10^{-6}$ & $1.3\times10^{-5}$\\
$230$ & H & $7.0\times10^{-5}$ & $7.0\times10^{-5}$ & $6.8\times10^{-5}$ & $6.6\times10^{-5}$ & $6.4\times10^{-5}$\\
      & D & $5.4\times10^{-5}$ & $5.2\times10^{-5}$ & $5.2\times10^{-5}$ & $5.0\times10^{-5}$ & $4.8\times10^{-5}$\\
\hline
\end{tabular}
\end{table}

\subsection{Least-squares fits}\label{app:chi2}

All fits of this work minimize
\begin{equation}
\chi^2\;=\;\sum_{i=1}^{n}\biggl[\frac{y_i-y_i^{\rm calc}}{\sigma_i}\biggr]^2 ,
\label{eq:chi2}
\end{equation}
where $y_i$ are the $n$ fitted values, $\sigma_i$ their errors and $y_i^{\rm calc}$ the
calculated values, which depend on the parameters. The number of degrees of freedom is
$n$ minus the number of parameters. If the model is correct and the errors are Gaussian
and correctly estimated, $\chi^2$ obeys the chi-square distribution for this number of
degrees of freedom. Its mean equals the number of degrees of freedom, and its standard
deviation is the square root of twice this number. Let $P$ be the probability that this
distribution exceeds the value of $\chi^2$ found. A small $P$ means that the deviations
are too large to come from the errors alone, and a fit with $P$ below $0.05$ is rejected.
A value of $P$ close to one means instead that the errors are overestimated. The errors
of the parameters are obtained from the curvature of $\chi^2$ at its minimum, through the
covariance matrix of the parameters. Correlations between parameters are included in the
errors of derived quantities, such as $s_p=e_p/T^*$.

The fits differ in the values $y_i$ and errors $\sigma_i$. For the coverage series the
values are the five published $E_d$ and the five $\ln\nu_0$ of Ref.~\cite{Graham1999}.
The error of $\ln\nu_0$ is the relative error of $\nu_0$. The parameters are the two
Arrhenius parameters of $\Gamma_0$ and $e_p$ and $T^*$, with $6$ degrees of freedom, or $7$
for the constant coupling. For the measured widths of the coverage series the values are the
logarithms of the $31$ widths, with the assumed relative errors of Sec.~5.2 and two
Arrhenius parameters, with $29$ degrees of freedom. The joint fit adds the two tilts at $0.15$ and
$0.2$~ML to the coverage series, with their errors from Table~\ref{tab:kfit}. This gives
$12$ values and $8$ degrees of freedom. Each momentum-resolved curve is fitted with \eqref{eq:shape}, linear
in $\bar\alpha$ and $\bar\alpha\delta$, with the digitized error bars as $\sigma_i$. For the friction of the substrate the values are the logarithms of the extracted HSE hop
rates, the published rates divided by $g_0$, with the published error bars divided in the
same way. The friction $\gamma$ is the only parameter. Its error is then scaled by the
square root of $\chi^2$ divided by the number of degrees of
freedom, when this ratio exceeds one. The common tilt of the four HSE curves,
$\delta=-0.40\pm0.04$, is the weighted mean of their tilts. The law $\delta\propto1/T$ is
fitted to the eight measured tilts at $0.1$~ML, with one parameter and $7$ degrees of
freedom. The closed form \eqref{eq:Rcf} is fitted to the $46$ simulated values of
$\lambda_{\min}/\Omega$ without weights, and the error of $\kappa$ is obtained from the
scatter of the residuals.

\subsection{Digitization of the data and robustness of the tilt}\label{app:digit}

All momentum-resolved data were digitized from the published figures at $200$--$400$
dots per inch. The HSE dephasing rates of Ref.~\cite{Jardine2010}, Fig.~2, are drawn in
colour. The error bars of each point were located as vertical lines of the colour of
the series, and the point was placed at the midpoint of its bar. Points with
overlapping bars were placed by hand. The fitted lines of the same figure were digitized
beyond the last data point. They are exact Chudley--Elliott forms with heights
$142.5$, $28.5$, $14.3$ and $3.12$~ns$^{-1}$ for H at $220$ and $140$~K, D at $140$~K
and H at $90$~K.

The QHAS widths of Ref.~\cite{Graham1999}, Figs.~7 and 8, are drawn in black, with
different symbols for each series. Each symbol was first located by eye and then
refined to the maximum of the correlation with a template of its shape. The axes were
calibrated on the tick labels. Each error bar was traced from the edge of its symbol
outwards, along the vertical line, up to its first cap. A cap is a horizontal segment
at least $14$ pixels wide. Where no bar extends beyond the symbol, the error was set to
the half size of the symbol. The median relative errors are $6$, $11$ and $9$\% for the
curves at $250$, $200$ and $160$~K on $\az{1\bar10}$, and $10$, $13$, $9$ and $16$\% for
the curves at $250$, $200$, $170$ and $140$~K on $\az{11\bar2}$. Where bars of different
series overlap, a bar may be attributed partly to another series. This can only make an
error too large. The digitization adds a scatter of a few per cent to the positions,
which raises the $\chi^2$ of the QHAS fits. The QHAS tilt errors are therefore likely
somewhat underestimated.

\label{app:robust}

The excess tilt of the HSE curves does not depend on a few points. The ratio of the
measured rates to the published Chudley--Elliott lines of Ref.~\cite{Jardine2010},
averaged with the error weights, is $1.23$--$1.42$ below $0.6$~\AA$^{-1}$, $1.01$--$1.12$
between $0.6$ and $1.0$~\AA$^{-1}$ and $0.92$--$0.97$ above $1.0$~\AA$^{-1}$, for the four
curves. The tilt is $-0.36$ to $-0.42$ with all points, $-0.31$ to $-0.41$ without the two
lowest-$K$ points, and $-0.38$ to $-0.48$ with the points up to $\bar{\rm M}$ only, with
errors of $0.05$--$0.16$. The largest change is that of H at $220$~K, from $-0.40$ to
$-0.31$ without the two lowest-$K$ points. These ratios refer to the published lines,
whereas Fig.~\ref{fig:shape}(a) uses the node rates $\bar\alpha$ fitted with the tilt.

A constant offset added to the Chudley--Elliott form fits the HSE curves as well as the
tilt. The offset is $0.11$, $0.18$, $0.17$ and $0.38$ times $\bar\alpha$ for H at $220$ and
$140$~K, D at $140$~K and H at $90$~K.

\subsection{Jumps through the hcp sites}\label{app:hcp}

Let each fcc site be linked to its three hcp neighbours, at the distance
$a/\sqrt3$, and each hcp site to its three fcc neighbours. An adsorbate leaves an fcc site
for each hcp neighbour at the rate $k_1$ and an hcp site for each fcc neighbour at the
rate $k_2$. In the dilute limit the Fourier transforms of the occupation probabilities
of the two sublattices obey a $2\times2$ rate equation. Its matrix has the diagonal
elements $3k_1$ and $3k_2$ and the off-diagonal elements $-3k_2\chi_{\K}$ and
$-3k_1\chi_{\K}^*$, with $\chi_{\K}=\tfrac13\sum_{\mathbf d}e^{i\K\cdot\mathbf d}$
the lattice sum over the three fcc--hcp vectors $\mathbf d$. The differences of these
vectors are the NN vectors of the fcc lattice, so that $|\chi_{\K}|^2=(1+2f_{\K})/3$.
The two eigenvalues are
\begin{equation}
\Lambda_\mp\;=\;\tfrac32(k_1+k_2)\Bigl\{1\mp\bigl[1-\tfrac83x(1-x)(1-f_{\K})\bigr]^{1/2}\Bigr\} ,
\qquad
x\;=\;\frac{k_1}{k_1+k_2} .
\end{equation}
The slow eigenvalue $\Lambda_-$ governs the decay at long times. When the hcp sites are
rarely occupied, $x\to0$ and $\Lambda_-\to2k_1(1-f_{\K})$. This is the
Chudley--Elliott form \eqref{eq:ce} of NN jumps between fcc sites, with the total hop
rate $\Gamma_0=2k_1$. When the hcp sites are populated, the next term of the expansion gives
$\Lambda_-/[1-f_{\K}]\propto1+\tfrac23x(1-x)(1-f_{\K})$. The ratio grows towards the
zone boundary, where $1-f_{\K}$ is largest. At $\bar{\rm K}$ it exceeds its
long-wavelength value by $5$\% for $x=0.05$ and by $25$\% for $x=0.2$. The measured
ratio grows towards long wavelength instead.

\subsection{Parameters and limits of the standard quantum TST}\label{app:tst}

With Ohmic friction the Wolynes factor \eqref{eq:Xi0} becomes a product over the
Matsubara frequencies $\mu_n=2\pi n/\beta\hbar$~\cite{Wolynes1981},
\begin{equation}
\Xi\;=\;\prod_{n=1}^{\infty}\frac{\omega_0^2+\mu_n^2+\mu_n\gamma}{-\omega^{\ddagger2}+\mu_n^2+\mu_n\gamma} .
\label{eq:Xi}
\end{equation}
It reduces to \eqref{eq:Xi0} at $\gamma=0$ and diverges when $\mu_1=\lambda^\ddagger$.

A fit of the HSE data with $n_{\rm dir}=3$ directions of escape required $\gamma\approx1000$~ps$^{-1}$~\cite{Jardine2010}.
The Grabert--Weiss theory gave about $14$ and $7$~ps$^{-1}$ for H and
D~\cite{Sanz2013,WeissGrabert1985}. The classical limit of \eqref{eq:tst} at high
temperature gives $V^\ddagger=72$~meV~\cite{TorresMiyares2025}, with a well frequency of
$30$~meV from vibrational spectroscopy. The cosine frequencies of \eqref{eq:parab}, used
in this work, lie about $8$\% below this value. They lower the TST prefactor by about
$15$\% above $150$~K and leave the slope unchanged. The same analysis narrows the
friction to $140$ and $70$~ps$^{-1}$~\cite{TorresMiyares2025}. The stochastic-wave-function
solutions with these values find the hops thermally activated to $90$--$100$\% at
$214$~K and to about $30$\% at $121$~K.

For a harmonic barrier coupled linearly to an Ohmic bath, \eqref{eq:tst} with
\eqref{eq:Xi} results from a normal-mode analysis. With $\hbar\gamma=92$~meV and
$\hbar\omega^\ddagger=27.7$~meV for H, $\lambda^\ddagger\approx\omega^{\ddagger2}/\gamma$,
and the crossover temperature $\hbar\lambda^\ddagger/2\pi k_B$ is about $14$~K for H and
$13$~K for D. Without friction it would not exceed $51$ and $36$~K. The measured
Arrhenius plots bend below about $110$~K, and the crossover was estimated at $66$ and
$63$~K~\cite{Jardine2010}. A crossover near $65$~K would require
$\hbar\lambda^\ddagger\approx36$~meV, that is, $\hbar\omega^\ddagger\approx68$~meV with
this friction. A fit of \eqref{eq:tst} to the HSE data down to $80$~K with
$\omega^\ddagger$ free indeed returns a barrier frequency about five times the well
frequency and $\gamma\approx1000$~ps$^{-1}$, the parameters of Ref.~\cite{Jardine2010}.

\subsection{Rate of the adlayer with neighbour-dependent barriers}\label{app:layer}

The occupied neighbours lower and soften the barrier of a jump. With the cosine
corrugation, a jump with $\nu$ occupied NN sites of the departing adparticle, other than
the target, has the barrier $V^\ddagger-e_p\nu$, and its frequencies are scaled by
\eqref{eq:parab}. Only the pair energy lowers the barrier here. The pair entropy enters the
Boltzmann factor through $\varepsilon(T)$. The factor
$B(\nu)=[\lambda^\ddagger\Xi/\omega^\ddagger]_\nu/[\lambda^\ddagger\Xi/\omega^\ddagger]_0$ is
the change of the dynamical and Wolynes factors of such a jump. The static factor of
\eqref{eq:layertst} then becomes
\begin{equation}
\Pi(\K)\;=\;\frac{(1-\theta)(1-\eta)\,\bigl\langle e^{-\beta\varepsilon\nu}B(\nu)\bigr\rangle_\nu}{F(\K)} ,
\label{eq:PiB}
\end{equation}
where the average $\langle\cdot\rangle_\nu$ runs over the binomial distribution of $\nu$ in
the pair approximation, over the $z-1$ such sites, each occupied with the probability $q$.
With $B=1$, \eqref{eq:PiB} reduces to $\Pi=\Gamma_{\rm eff}/\Gamma_0F$ with $\Gamma_{\rm eff}$
of \eqref{eq:gc}. The values of Sec.~6.4 are calculated with \eqref{eq:PiB}.

For H at $100$~K, averaged over the neighbourhoods of a jumping adsorbate, the barrier is
$65.5$ and $47.4$~meV at $0.1$ and $0.4$~ML, against $72.0$~meV for the isolated adsorbate.
The factors $\lambda^\ddagger/\omega^\ddagger$ and $\Xi$ are lowered accordingly. A denser
layer is thus more classical. Its calculated isotope ratio is lower by $1$--$2$\% at
$0.1$~ML and by $3$--$6$\% at $0.4$~ML, a prediction for a measurement on both isotopes.
Three cautions apply. The factorization into adsorbate and layer is exact only for $B=1$.
Otherwise $B(\nu)$ couples the adsorbate and the layer. The TST factors hold only above the
crossover. Away from the momentum transfer of the coverage series, $\lambda_{\min}/\Omega$
has no closed form and is simulated.

\subsection{The action of the barrier and the spectral action}\label{app:action}

The action of the barrier in the TST of one adsorbate has a counterpart in the adlayer. The weights of the relaxation modes in \eqref{eq:modes} can be
written as exponentials of a spectral action, a function of the relaxation rate. It
corresponds to the Euclidean action of the barrier~\cite{Weiss1999}, with the energy in the
role of the relaxation rate and the imaginary time $\hbar\beta$ in the role of $t$. For a
parabolic barrier the action is linear in the energy. The anharmonicity of a real barrier
makes it curve, and this curvature is its leading correction. The ideal layer has a single
rate, the counterpart of the parabolic barrier. The spread of the rates, the collective
friction, is then the counterpart of the curvature of the action produced by the
anharmonicity of the barrier.

\subsection{Scope of the analysis}\label{app:scope}

The analysis is well founded above about $130$~K, in the QHAS range. There the memory equation, the
statics and the hop rate of the isolated adsorbate all rest on established ground. Between
$130$ and $80$~K the coupling and the jump rates \eqref{eq:rate} are extrapolated, and the
interaction factor $g$ at $0.1$~ML lies between $0.97$ and $1.17$, depending on the jump rule
(Table~\ref{tab:rule}). Below $80$~K no data exist.
The tunnelling remains incoherent over the whole range (Sec.~6.3). An occupied neighbour
also shifts the energy of a well by $e_p\nu$, $10$--$40$~meV, far more than a tunnelling
bandwidth, so that the coverage alone would destroy coherent tunnelling.

Four limitations apply throughout. All data are digitized from published figures
(Appendix~\ref{app:digit}), and the QHAS widths of the coverage series carry no
published errors. The independent test of the coupling rests on one momentum-resolved curve at each of
$0.15$ and $0.2$~ML, with digitized errors that may be underestimated. Below about $130$~K the interaction factor $g$ depends
on the jump rates \eqref{eq:rate}, which no theory of incoherent tunnelling fixes in a
crowded environment. The published rates come from single-exponential fits of decays
that are not available, so that the collective friction is obtained from simulations
with the fitted coupling, not directly from the data.

\section*{Acknowledgements}

This work was supported by the Ministry of Science, Innovation and Universities
(Spain) under Grant No.~PID2023-149406NB-I00.

\section*{Data availability}

The data extracted from Refs.~\cite{Graham1999,Jardine2010} and the codes for the
closed forms and the kinetic Monte Carlo simulations are available from the
author on request.

\end{document}